\documentclass[12pt]{article}

\usepackage{amsmath}
\usepackage{amssymb}
\usepackage{amsthm}
\usepackage{mathrsfs}
\usepackage[T1]{fontenc}
\usepackage{enumerate}
\usepackage{color}

\def\1loop{conformal invariance conditions}

\def\eqn{equation}

\def\comrel{comutation relation}
\def\tfn{transformation}

\def\sm{sigma model}
\def\pl{Poisson-Lie}

\def\dd{Drinfel'd double}
\def\sugras{Generalized Supergravity Equations\ }
\def\vbe{beta function equations}
\def\4diml{four-dimensional}
\def\bkg{background}
\def\wrt{with respect to}
\def\-1{^{-1}}
\def\half{\frac{1}{2}}
\def\coor{coordinate}

\def\ca{{\mathfrak a}}
\def\cd{{\mathfrak d}}
\def\cg{{\mathfrak g}}
\def\tcg{\tilde{\mathfrak g}}
\def\hcg{\hat{\mathfrak g}}
\def\bcg{\bar{\mathfrak g}}

\def\wt{\tilde}
\def\wh{\widehat}
\def\wwt{\widetilde}
\def\sm{sigma model}
\def\pltp{Poisson--Lie T-pluralit}
\def\pltd{Poisson--Lie dualit}
\def\plti{Poisson--Lie identit}
\def\natd{(non-)Abelian T-dualit}

\def\cf{{\mathcal {F}}}
\def\ci{{\mathcal {J}}}

\newcommand{\unit}{\mathbf{1}}
\newcommand{\nul}{\mathbf{0}}
\newcommand{\A}{\mathscr{A}}
\newcommand{\B}{\mathscr{B}}
\newcommand{\M}{\mathscr{M}}
\newcommand{\D}{\mathscr{D}}
\newcommand{\G}{\mathscr{G}}
\newcommand{\tG}{\widetilde{\mathscr{G}}}
\newcommand{\hG}{\widehat{\mathscr{G}}}
\newcommand{\bG}{\bar{\mathscr{G}}}

\newcommand{\N}{\mathscr{N}}

\newcommand{\PL}{Poisson--Lie}
\newcommand{\Bsh}{B-shift}
\newcommand{\betash}{$\beta$-shift}
\newcommand{\wb}{\bar}

\title{Poisson--Lie identities and dualities of Bianchi cosmologies}
\author{Ladislav Hlavat\'y\footnote{hlavaty@fjfi.cvut.cz}
\\ {\em Faculty of Nuclear Sciences and Physical Engineering,}
\\ {\em Czech Technical University in Prague,}
\\ {\em Czech Republic}
\and
Ivo Petr\footnote{ivo.petr@fit.cvut.cz}
\\ {\em Faculty of Information Technology,}
\\ {\em Czech Technical University in Prague,}
\\ {\em Czech Republic}}

\begin{document}
\maketitle

\begin{abstract}
We investigate a special class of Poisson--Lie T-plurality transformations of Bianchi cosmologies invariant with respect to non-semisimple Bianchi groups. For six-dimensional semi-Abelian Manin triples $\mathfrak{b}\bowtie\mathfrak{a}$ containing Bianchi algebras $\mathfrak{b}$ we identify general forms of Poisson--Lie identities and dualities. We show that these can be decomposed into simple factors, namely automorphisms of Manin triples, B-shifts, $\beta$-shifts, and ``full'' or ``factorized'' dualities. Further, we study effects of these transformations and utilize the decompositions to obtain new backgrounds which, supported by corresponding dilatons, satisfy Generalized Supergravity Equations.
\end{abstract}


\tableofcontents


\section{Introduction}

Duality transformations play crucial role in the study of
various aspects of string theory and related fields. They connect
theories in different coupling regimes or, in the case of
T-duality, backgrounds with distinct curvature properties. Both
Abelian T-duality \cite{buscher:ssbfe} and its non-Abelian
generalization \cite{delaossa:1992vc,rocekver:duality} rely on the
presence of symmetries of \sm\ \bkg s. Dual \sm\ related to the
original one by T-duality is obtained by gauging of the symmetry
and introduction of Lagrange multipliers. However, the symmetries
are not preserved in the non-Abelian case, meaning we may not be
able to return to the original model by dualization. Despite this
serious issue we see renewed interest in \natd y (NATD). The
procedure was extended to RR fields in \cite{1012.1320
sfethomp,1104.5196LCST} and is used frequently to find new
supergravity solutions, see e.g. \cite{KlebWitt, INNSZ} and
references therein. It also applies in the study of integrable
models \cite{BorWulff1, hoatsey, BorWulff}.  
In this paper we are going to present new solutions of the \sugras obtained by  action of Poisson-Lie
identities and \pltd ies introduced in \cite{hlape:pltprev} on the Bianchi cosmologies  \cite{Batakis}. This extends list of results obtained in \cite{hokico}.

\PL\ T-duality \cite{klise} introduces \dd\ as the underlying algebraic
structure of T-duality and replaces symmetry of the \sm\ \bkg\ by
the so-called \PL\ symmetry \cite{klim:proc}. This allows us to
treat both models equally and solves the above mentioned problem. We
shall use this formalism through the whole paper. In the case of
\natd y the Lie group $\D$ of \dd\ splits into Lie subgroups $\G$
and $\tG$ of equal dimension, where the former represents symmetries
of the original \bkg\ while the latter is Abelian. In this paper we
consider only these semi-Abelian \dd s as we focus on dualization of
particular \bkg s and the presence of symmetries remains crucial in
such case\footnote{See \cite{Bouwknegt2017} for discussion on this
topic.}. \pltd y exchanges roles of $\G$ and $\tG$, and we
understand it as a change of decomposition $(\G|\tG)$ of $\D$ to
$(\tG|\G)$ and vice versa. Beside $(\G|\tG)$ and $(\tG|\G)$ there
might be other decompositions $(\hG|\bG),(\bG|\hG)$ of a \dd\ $\D$
that can be used to construct mutually related \sm s. The
corresponding transformation between \sm s was denoted \pltp y
\cite{unge:pltp}. Decompositions of low-dimensional \dd s were
classified in papers \cite{gom:ctd, hlasno:pltdm2dt,
snohla:ddoubles} in terms of Manin triples $(\cd,\cg,\tcg)$ that
represent decompositions of Lie algebra $\cd$ of the \dd\ $\D$ into
subalgebras $\cg$ and $\tcg$ corresponding to subgroups $\G$ and
$\tG$.

In our recent paper \cite{hlape:pltprev} we noted that besides
\natd y there exist other \tfn s that either preserve or exchange
the algebras $\mathfrak g$ and $\wt{\mathfrak g}$ of the Manin
triple $(\cd,\cg,\tcg)$. We shall call them \emph{Poisson-Lie
identities} and \emph{\pltd ies}. Similar transformations were
studied in \cite{LuOst} to get insight into the structure of the
so-called NATD group of T-duality transformations. Beside others,
this group contains automorphisms of the algebras forming Manin
triples, B-shifts, $\beta$-shifts\footnote{$\beta$-shifts are also referred to as TsT \tfn s, see \cite {frolov,ostenger}.}, ``factorized'' dualities and
their compositions. These, however, have to be understood as special
cases of \pltp y. We continue the investigation of the NATD group
probing its structure for low-dimensional \dd s, where general forms
of \PL\ identities and dualities can be identified. 
We show that all transformations are 
finite compositions
of the special elements of NATD group that were mentioned earlier.
It turns out that the effect of automorphisms and B-shifts on 
resulting \bkg s can be often eliminated by a change of coordinates,
hence, we identify what parameters of the
transformations are relevant.

Long-lasting problem appearing in discussion of non-Abelian
T-duality is that dualization with respect to non-semisimple group
$\G$ leads to mixed gauge and gravitational anomaly, see
\cite{aagl}, proportional to the trace of structure constants of
$\cg$. Authors of paper \cite{hokico} have found non-Abelian T-duals
of Bianchi cosmologies \cite{Batakis} and have shown that instead of
standard \vbe\ the dual backgrounds satisfy the so-called
Generalised Supergravity Equations containing Killing vector $\ci$ whose
components are given by the trace of structure constants. For Bianchi $V$ cosmology this was observed already in \cite{saka1}. Therefore,
it is natural to ask if \bkg s and dilatons obtained from
Bianchi cosmologies by \PL\ identities and dualities satisfy
Generalised Supergravity Equations as well and what Killing vectors
have to be used. We show that components of $\ci$ are determined by trace of structure constants, however, one must carefully inspect what groups (or subgroups of) $\G$ truly participate in the duality transformation. For Yang--Baxter deformations of $AdS_5\times S^5$ the problem of finding $\ci$ was addressed in \cite{ABCSSY1,ABCSSY2} and physical meaning of $\ci$ was found in \cite{ACSSY}. Discussion on the \pltd y and plurality of the Generalized Supergravity Equations can be found e.g. in \cite{dehato} and \cite{saka2}.

We start with a short description of \pltp y in section
\ref{basics_of_pltp}, where necessary formulas are summarized
and general forms of transformed \bkg s are presented. In sections
\ref{B_V}--\ref{B_II} we investigate various  transformations
of Bianchi cosmologies
 focusing on groups that are not
semisimple. Since calculations with general transformations often result
in rather complicated backgrounds that cannot be displayed, detailed
description is given only for special elements of the NATD group.
Summary of transformed \bkg s can be found in the Appendix.

\section{Basics of \pltp y}\label{basics_of_pltp}

In the first two subsections we recapitulate \pltp y with spectators \cite{klise,unge:pltp,hlapevoj}. We follow the summary given in \cite{hlape:pltprev}.

\subsection{Sigma models}

Let $\M$  be $(n+d)$-dimensional (pseudo-)Riemannian target manifold and consider sigma model on $\M$ given by Lagrangian
\begin{equation}\nonumber 
{\cal L}=\partial_- \phi^{\mu}\cf_{\mu\nu}(\phi)\partial_+ \phi^\nu,\qquad
\phi^\mu=\phi^\mu(\sigma_+,\sigma_-), \qquad
\mu=1,\ldots,n+d
\end{equation}
where tensor field $\cf=\mathcal G + \mathcal B$ on $\M$ defines
metric and torsion potential (Kalb--Ramond field) of the target
manifold. Assume that there is a $d$-dimensional Lie group $\G$ with
free action on $\M$ that leaves the tensor invariant. The action of
$\G$ is transitive on its orbits, hence we may locally consider
$\M\approx (\M/\G) \times \G = \N \times \G$, and introduce adapted
coordinates
\begin{equation}
\nonumber
\{s_\alpha,x_a\},\qquad \alpha=1, \ldots,n = \dim \N ,\ \ a=1,
\ldots, d = \dim \G
\end{equation}
where $s_\alpha$ label the orbits of $\G$ and are treated as
spectators, and $x_a$ are group coordinates\footnote{Detailed
discussion of the process of finding adapted coordinates can be
found e.g. in \cite{hlapevoj,hlafilp:uniq,php}}. Dualizable sigma
model on $\N \times \G$ is given by tensor field $\mathcal{F}$
defined by $(n+d)\times (n+d)$ matrix $E(s)$ as
\begin{equation}\label{F}
\cf(s,x)=\mathcal{E}(x)\cdot E(s)\cdot \mathcal{E}^T(x), \qquad
\mathcal{E}(x)=
\left(
\begin{array}{cc}
 \unit_n & 0 \\
 0 & e(x)
\end{array}
\right)
\end{equation}
where $e(x)$ is $d\times d$ matrix of components of right-invariant Maurer--Cartan form $(dg)g^{-1}$  on $\G$.

Using non-Abelian T-duality one can find dual sigma model on
{$\N\times \A$, where $\A$ is Abelian subgroup of semi-Abelian \dd\
$\D=(\G|\A)$.} The necessary formulas will be given in the following
subsection as a special case of \pltp y.  In this paper the groups
$\G$ will be non-semisimple Bianchi groups.
Bianchi cosmologies are defined on four-dimensional manifolds, hence $d=3$, $n=1$, and we denote the spectator as $t:=s_1$. Elements of the group $\G$ shall be parametrized as $g=e^{x_1 T_1}e^{x_2 T_2}e^{x_3 T_3}$ where $e^{x_2 T_2}e^{x_3 T_3}$
and $e^{x_3 T_3}$ are normal subgroups of $\G$. Similarly, elements of $\A$ are parametrized as $\wt g=e^{\wt x_1 \wwt T^1}e^{\wt x_2 \wwt T^2}e^{\wt x_3 \wwt T^3}$.

\subsection{\PL\ T-plurality with spectators}\label{pltpuvod1}

For certain \dd s several decompositions may exist. Suppose that $\D=(\G|\tG)$ splits into another pair of subgroups $\hG$ and $\bG$. Then we can apply the full framework of \PL\ T-plurality \cite{klise, unge:pltp} and find sigma model on $\N \times \hG$.

The $2d$-dimensional Lie algebra $\cd$ of the \dd\ $\D$ is equipped
with an ad-invariant non-degenerate symmetric bilinear form $\langle
. , . \rangle$. Let $\cd=\cg \bowtie \tcg$ and $\cd=\hcg \bowtie
\bcg$ be two decompositions (Manin triples $(\cd, \cg, \tcg)$ and
$(\cd,\hcg, \bcg)$) of $\cd$ into double cross sum of
subalgebras \cite{majid}
 that are maximally isotropic with respect to $\langle . , .
\rangle$. The pairs of mutually dual bases $T_a \in \cg,\
\widetilde{T}^a \in \tcg$ and $\wh T_a \in \hcg,\ \wb{T}^a \in
\hcg$, $a=1, \ldots, d,$ satisfying
\begin{align}
\label{dual_bases}
\langle T_a, T_b \rangle & = 0, & \langle \widetilde{T}^a, \widetilde{T}^b \rangle &= 0, & \langle T_a, \widetilde{T}^b \rangle & = \delta_a^b,\\
\nonumber \langle \wh T_a, \wh T_b \rangle & = 0, & \langle \wb{T}^a, \wb{T}^b \rangle & = 0, & \langle \wh T_a, \wb{T}^b \rangle & =  \delta_a^b
\end{align}
then must be related by transformation
\begin{equation}\label{C_mat}
\begin{pmatrix}
\wh T \\
\wb T
\end{pmatrix}
 = C \cdot
\begin{pmatrix}
T \\
\widetilde T
\end{pmatrix}
\end{equation}
where $C$ is an invertible $2d\times 2d$ matrix. Due to ad-invariance of the bilinear form $\langle . , . \rangle$ the algebraic structure of $\mathfrak{d}$ is given both by
\begin{equation}
\label{commutation_on_d}
[T_i,T_j]=f_{ij}^k T_k, \qquad [\widetilde T^i, \widetilde T^j]=\wt f^{ij}_k \widetilde T^k, \qquad [T_i,\widetilde T^j]=f_{ki}^j \widetilde T^k + \wt f_i^{jk} T_k
\end{equation}
and
\begin{equation}
\label{commutation_on_d2}
[\wh T_i, \wh T_j]=\hat f_{ij}^k \wh T_k, \qquad [\wb T^i, \wb T^j]=\wb f^{ij}_k \wb T^k, \qquad [\wh T_i,\wb T^j]=\hat f_{ki}^j \wb T^k + \wb f_i^{jk} \wh T_k.
\end{equation}
Given the structure constants $F_{ij}^k$ of $\cd=\cg \bowtie \tcg$ and $\wh F_{ij}^k$ of $\cd=\hcg \bowtie \bcg$, the matrix $C$ has to satisfy equation
\begin{equation}
\label{C_str_const}
C_a^p C_b^q F_{pq}^r = \wh F_{ab}^c C_c^r.
\end{equation}
To preserve the bilinear form $\langle . , . \rangle$ and thus \eqref{dual_bases}, $C$ also has to satisfy
\begin{equation}
\label{C_bilin}
C_a^p C_b^q (D_0)_{pq}=(D_0)_{ab}
\end{equation}
where $(D_0)_{ab}$ are components of matrix $D_0$ that can be
written in block form as
\begin{equation}
\label{Delta} D_0=\begin{pmatrix} \mathbf{0}_d & \mathbf{1}_d \\
\mathbf{1}_d & \mathbf{0}_d \end{pmatrix}.
\end{equation}
In other words, $C$ is an element of $O(d,d)$ but, unlike the case of Abelian T-duality, not every element of $O(d,d)$ is allowed in \eqref{C_mat}.

For the following formulas it will be convenient to introduce $d \times d$ matrices $P, Q, R, S$ as
\begin{equation}\label{pqrs}
\begin{pmatrix}
T \\
\widetilde T
\end{pmatrix}
= C^{-1} \cdot
\begin{pmatrix}
\wh T \\
\wb T
\end{pmatrix} =
\begin{pmatrix}
 P & Q \\
 R & S
\end{pmatrix} \cdot
\begin{pmatrix}
\wh T \\
\wb T
\end{pmatrix}
\end{equation}
and extend these to $(n+d)\times (n+d)$ matrices
\begin{equation}
\nonumber
\mathcal{P} =\begin{pmatrix}\unit_n &0 \\ 0&P \end{pmatrix}, \qquad \mathcal{Q} =\begin{pmatrix}\nul_n&0 \\ 0&Q \end{pmatrix}, \qquad \mathcal{R} =\begin{pmatrix}\nul_n&0 \\ 0&R \end{pmatrix}, \qquad \mathcal{S} =\begin{pmatrix}\unit_n &0 \\ 0& S \end{pmatrix}
\end{equation}
{to accommodate the spectator fields.} It is also advantageous to
introduce block form of $E(s)$ as
\begin{equation}
\nonumber
 E(s)=
\left(
\begin{array}{cc}
 E_{\alpha\beta}(s) & E_{\alpha b}(s) \\
 E_{a\beta}(s)      & E_{ab}(s)
\end{array}
\right), \qquad
\alpha, \beta=1,\ldots,n, \quad a, b=1, \ldots, d.
\end{equation}
The sigma model on $\N \times \hG$ related to \eqref{F} via \pltp y
is given by tensor
\begin{equation} \label{Fhat} \widehat{\cf}(s,\hat x)=
\mathcal{\widehat E}(\hat x)\cdot \widehat E(s,\hat x) \cdot
\mathcal{\widehat E}^T(\hat x), \qquad \mathcal{\widehat E}(\hat x)=
\begin{pmatrix}
\unit_n & 0 \\
 0 & \wh e(\hat x)
\end{pmatrix}
\end{equation}
where $\wh e(\hat x)$ is $d\times d$ matrix of components of
right-invariant Maurer--Cartan form $(d\hat g)\hat g^{-1}$  on
$\wh\G$,
\begin{equation}\label{Fhat2} \wh
E(s,\hat x)=(\unit_{n+d}+\wh E(s) \cdot \wh{\Pi}(\hat x))^{-1}\cdot
\wh E(s) { =(\wh E\-1(s)+ \wh{\Pi}(\hat x))^{-1}},
\end{equation}
$$\widehat\Pi(\hat x)= \left(
\begin{array}{cc}
\nul_n & 0 \\
 0 & \widehat b(\hat x) \cdot \widehat a^{-1}(\hat x)
\end{array}
\right),$$
and matrices $\widehat b(\hat x)$ and $\widehat a(\hat x)$ are
submatrices of the adjoint representation $$ ad_{{\hat g}\-1}(\wh
T) = \widehat b(\hat x) \cdot \wb T + \widehat a^{-1}(\hat
x)\cdot\wh T.
$$
The matrix $\wh E(s)$ is obtained by formula
\begin{equation}\label{E0hat}
\wh E(s)=(\mathcal{P}+ E(s) \cdot \mathcal{R})^{-1} \cdot
(\mathcal{Q}+E(s) \cdot \mathcal{S})
\end{equation}
so it is necessary\footnote{{Invertibility of $\wh E(s)$ is not required in the first expression in \eqref{Fhat2} and only $\det\, \left(\mathcal P+E(s)\cdot \mathcal R \right)\neq 0$ is required. However, for regular $\wh E (s)$ the formulas simplify.}} that
\begin{equation}\nonumber 
\det\, \left(\mathcal P+E(s)\cdot \mathcal R \right)
{\neq 0 \neq \det\, \left(\mathcal{Q}+E(s) \cdot \mathcal{S} \right)}.
\end{equation}
Formulas \eqref{Fhat}--\eqref{E0hat} reduce to those for full \pltd
y if we choose $P=S=\mathbf{0}_d$ and $Q=R=\mathbf{1}_d$.
Furthermore, for a semi-Abelian \dd\ the well-known Buscher rules
for \natd y are restored. If there are no spectators the plurality
is called atomic.

\subsection{\PL\ identities and \pltd ies}\label{factors}

Let us now restrict our considerations to mappings \eqref{C_mat}
that preserve the Manin triple, i.e. $\hcg = \cg,\, \bcg = \tcg,$ and
that satisfy \eqref{C_str_const} and \eqref{C_bilin}. They are the
\plti ies. The Maurer--Cartan form $(dg)g^{-1}$ remains
unchanged but $E(s)$ transforms as in \eqref{E0hat}. Moreover, for
non-Abelian T-duality  the algebra
$\tcg$ is Abelian, i.e. $\tcg=\ca$, thus $\wh b$ and $\wh \Pi$
vanish, and we may write\footnote{Since $\hG=\G$ we omit the hat
over group coordinates and write simply $x$.}
\begin{equation}\nonumber 
\wh \cf(s,x)=\mathcal{E}(x)\cdot \wh E(s)\cdot \mathcal{E}^T(x).
\end{equation}
Let us note that both
\bkg s $\cf(s,x)$ and $\wh \cf(s,x)$ are invariant \wrt\  the group $\G$.

For special transformations mentioned in the introduction we can
further specify the resulting \bkg s. Namely,
matrices
\begin{equation}\label{IA}
I_A= \left(\begin{array}{cc}
 A  & 0  \\
 0  & A^{-T}  \\
\end{array}\right)
\end{equation}
are always among the transformations \eqref{C_mat} preserving the
Manin triple $\cd=\cg\bowtie \ca$ if $A$ is an
automorphism\footnote{Our approach differs from \cite{LuOst}, where
the authors consider vector space isomorphisms of Lie algebra of the \dd {}
rather than automorphisms of a chosen Manin triple.} of $\cg$. Transformed
$\wh E(s)$ then reads
\begin{equation}\nonumber 
\wh E(s)=\mathcal A\cdot E(s)\cdot \mathcal A^{T}, \qquad
\mathcal A=
\left(
\begin{array}{cc}
 \unit_n & 0 \\
 0 & A
\end{array}
\right).
\end{equation}

Transformations \eqref{C_mat} of the form
\begin{equation}\label{IB}
I_B=
\left(
\begin{array}{cc}
\unit_d & B  \\
 0  & \unit_d
\end{array}
\right), \qquad B^T=-B
\end{equation}
are called B-shifts, since  the \bkg\ $\wh \cf(s,x)$ obtained by this \tfn\ is  given by
\begin{equation}\nonumber 
\wh E(s)=\left(E(s)- \bar B\right), \qquad \bar B= \left(\begin{array}{cc} \nul_n & 0 \\ 0 & B\end{array}\right).
\end{equation}
Tensor $\wh \cf$ differs from the original one by an antisymmetric term  $\mathcal B' = - \mathcal{ E}(x)\cdot \bar B \cdot \mathcal{ E}(x)^T$
that, however, for solvable Bianchi algebras does not produce supplementary torsion. Therefore, for all investigated Bianchi cosmologies $\wh \cf$ is gauge equivalent to the initial
tensor $\cf$.

$\beta$-shifts are generated by \tfn\ matrices
\begin{equation}\nonumber 
I_\beta=
\left(
\begin{array}{cc}
 \unit_d  & 0  \\
 \beta  & \unit_d
\end{array}
\right), \qquad \beta^T=-\beta
\end{equation}
and the transformed $\wh \cf(s,x)$ is given by
\begin{equation}\nonumber 
\wh E(s)=\left(\unit -E(s) \cdot
\bar\beta\right)\-1\cdot E(s), \qquad
\bar\beta= \left(
\begin{array}{cc}
 \nul_n & 0 \\
 0 & \beta
\end{array}
\right)
.
\end{equation}
For invertible $E(s)$, we may write $\wh E(s)=\left(E(s)\-1 -\bar\beta \right)\-1$.

Beside these transformations we may also encounter mappings $I_F$
that switch some of the basis vectors $T_i \leftrightarrow \wwt T_i$
while preserving structure coefficients of the Manin triple.
These ``factorized'' dualities can be interpreted as dualization
with respect to subgroups of $\G$. In general, these cannot be
written concisely in block form and we do not discuss them here. We
shall see many examples in the following sections.

Let us further investigate \pltd ies, i.e. mappings \eqref{C_mat} that change Manin triple
$\cd=\cg\bowtie \ca$ to $\cd=\ca\bowtie \cg$.  Equation \eqref{C_mat} implies that
\begin{equation}\nonumber
\begin{pmatrix}
\wwt T \\
 T
\end{pmatrix}
 = D_0 \cdot
\begin{pmatrix}
T \\
\wwt T
\end{pmatrix}
 = D_0 \cdot I \cdot
\begin{pmatrix}
T \\
\wwt T
\end{pmatrix}.
\end{equation}
Therefore, \pltd ies are composed of \PL\
identities $I$ and ``full'' T-duality $D_0$ that exchanges all
generators of $\mathfrak g$ and $\ca$ as $T_i \leftrightarrow \wwt
T_i$ for $i= 1,\ldots,d$. In this way we can define dual B-shifts, dual
$\beta$-shifts and dual automorphisms.

Backgrounds on
$\tG=\A$ obtained from \eqref{F} by \pltd ies have the form
\begin{equation}\nonumber 
{\wwt\cf(s,\wt x)=\left(\wh E\-1(s)+\wwt \Pi(\wt x)\right)\-1}, \qquad
\wwt\Pi(\wt x)=
\left(
\begin{array}{cc}
 \nul_n & 0 \\
 0 & \wwt b (\wt x)
\end{array}
\right)
\end{equation}
because $\wwt e(\wt x)=\wwt a (\wt x)=\unit_d$. For solvable groups
$\G$ we have $\wwt b_{ab}(\wt x)=f_{ab}^c \, \wt x_c$.

It is possible to be more specific when we restrict to specific elements of the NATD group. In the presence of spectators, however, the formulas for \pltd ies are quite complicated. Fortunately, we do not need their full form since $E_{a\beta}(s)$ and $E_{\alpha b}(s)$ in $E(s)$ vanish for the \bkg s discussed in the rest of the paper. Hence $\wh \cf_{\alpha\beta}(s)=\cf_{\alpha\beta}(s)$ and plurality affects  only $\cf_{ab}(s,x)$. The transformations we are interested in, therefore, concern only the $\wh \cf_{ab}(s,\hat x)$ block of the resulting \bkg\ tensor $\wh \cf$.

For the dual B-shift
\begin{equation}\nonumber 
D_B=D_0 \cdot I_B=
\left(
\begin{array}{cc}
 0  & \unit_d \\
 \unit_d & B
\end{array}
\right), \quad B^T=-B
\end{equation}
and vanishing $E_{a\beta}(s)$, $E_{\alpha b}(s)$ the matrix $\left(\wh E\-1(s)+\wwt \Pi(\wt x)\right)$ has the form
\begin{equation}\nonumber
\left(
\begin{array}{cc}
E_{\alpha\beta}\-1(s) & 0 \\
0 & E_{ab}(s)-B_{ab}+\wwt b_{ab}(\wt x)
\end{array}
\right).
\end{equation}
For solvable groups $(\wwt b(\wt x)-B)_{ab}=f_{ab}^c \wt x_c - B_{ab}$. As we shall see, for some groups this enables us to get rid of some parameters of $B_{ab}$ by \coor\ \tfn s.

General formulas for dual $\beta$-shift
\begin{equation}\nonumber 
D_{\beta}=D_0 \cdot I_{\beta}=
\left(
\begin{array}{cc}
 \beta  & \unit_d \\
 \unit_d & 0
\end{array}
\right), \quad \beta^T=-\beta
\end{equation}
in the presence of spectators  are complicated and not particularly illuminating. For vanishing $E_{a\beta}(s)$, $E_{\alpha b}(s)$ one  gets
\begin{equation}\nonumber
\wh E(s)=
\left(
\begin{array}{cc}
 E_{\alpha\beta}(s)&0 \\
0 & E_{ab}\-1(s)-\beta
\end{array}
\right).
\end{equation}

Let us focus on the role of automorphisms $I_A$ and their duals now.
As conjectured in \cite{LuOst}, it turns out that all \PL\ identities and \pltd ies are generated by automorphisms of Manin triples, \Bsh s, \betash s and factorised dualities. Moreover, in most of the examples of Bianchi cosmologies discussed later we find that the general $C$ matrix in \eqref{C_mat} splits as
$$
C=I_{A_2}\cdot C' \cdot I_{A_1}
$$
where $C'$ is either $I_B$, $I_\beta$, $I_F$ or their duals.
Transformed \bkg s then can be written as
\begin{equation}\label{Fhat_A2}
\widehat{\cf}(s,\hat x)=
\mathcal{\widehat E}(\hat x)\cdot \mathcal A_2 \cdot \wh E'(s,\hat x)\cdot \mathcal A_2^T \cdot
\mathcal{\wh E}^T(\hat x)
\end{equation}
where
\begin{equation}\nonumber
\wh E'(s,\hat x) = \left(\left(\mathcal{Q}' +\mathcal A_1 E(s) \mathcal A_1^T \mathcal{S}'\right)\-1 \left(\mathcal P' +\mathcal A_1 E(s) \mathcal A_1^T  \mathcal R'\right)+ \mathcal A_2^T \wh\Pi(\hat x) \mathcal A_2 \right)\-1
\end{equation}
and $\mathcal{P}',\mathcal{Q}',\mathcal{R}',\mathcal{S}'$ are found from $C'$.
As expected, these expressions can be interpreted as application of \pltp y on a \bkg\ given by matrix $E'(s)=\mathcal A_1 E(s) \mathcal A_1^T$. Important is that $\mathcal A_2$ can be eliminated from \eqref{Fhat_A2} by suitable transformation of group coordinates. Using the transformation properties of the covariant tensor $\wh \cf$ we may try to integrate the Jacobi matrix
\begin{equation}\label{Jac_transf}
J^{\mu}_{\lambda}=\frac{\partial \hat x'^{\mu}}{\partial \hat x^{\lambda}} = \wh e(\hat x)_{\lambda}^{^k} (A_2)_{k}^{p} (\wh e\-1(\hat x))^{\mu}_{p}
\end{equation}
to find coordinates $\hat x'$ such that \eqref{Fhat_A2} simplifies to
$$
\widehat{\cf}(s,\hat x)=
\mathcal{\widehat E}(\hat x) \cdot \wh E'(s,\hat x)\cdot \mathcal{\wh E}^T(\hat x).
$$
These transformations can be always found for \pltd ies on semi-Abelian \dd\ where $\tcg=\ca$ and $\wh e(\hat x)=\unit$. In this case $J = A$, the transformation is linear and can be even combined with the coordinate shifts mentioned earlier for dual B-shifts.

\subsection{Generalized Supergravity Equations and transformation of dilaton}

One of the goals of this paper is to verify whether \bkg s obtained from \plti ies and dualities satisfy Generalized Supergravity Equations of Motion. 
We adopt convention used in \cite{hokico} so the equations read\footnote{We consider purely bosonic \bkg s.}
\begin{align}\label{betaG}
0&=R_{\mu\nu}-\frac{1}{4}H_{\mu\rho\sigma}H_{\nu}^{\
\rho\sigma}+\nabla_{\mu}X_{\nu}+\nabla_{\nu}X_{\mu},\\ \label{betaB}
0 &=
-\frac{1}{2}\nabla^{\rho}H_{\rho\mu\nu}+X^{\rho}H_{\rho\mu\nu}+\nabla_{\mu}X_{\nu}-\nabla_{\nu}X_{\mu},\\
\label{betaPhi} 0 &=
R-\frac{1}{12}H_{\rho\sigma\tau}H^{\rho\sigma\tau}+4\nabla_{\mu}X^{\mu}-4X_{\mu}X^{\mu}
\end{align}
where $\nabla$ is covariant derivative and
\begin{equation}\nonumber
X_{\mu}=\partial_{\mu}\Phi + \mathcal J^{\nu} \cf_{\nu\mu}.
\end{equation}
For vanishing vector $\mathcal{J}$ the usual one-loop beta function equations, i.e. \1loop, are recovered.

Under \natd y dilaton transforms as
\begin{equation}
\label{dualdilaton}
\wwt \Phi = \Phi + \frac{1}{2}\ln \det M
\end{equation}
where matrix $M$ is given by ``group block'' $E_{ab}$ of $E(s)$ and submatrices of adjoint representation as
\begin{equation}\nonumber
M=(E_{ab}(s)+\wwt b (\wt x)\cdot\wwt a^{-1} (\wt x))^{-1} =(\wwt E^{-1}_{ab}(s)+\wwt b (\wt x)\cdot\wwt a^{-1} (\wt x))^{-1}.
\end{equation}
The formula \eqref{dualdilaton} can be utilized not only for ``full'' duality given by $D_0$, but also for factorized dualities. However, for successful application of this rule it is necessary to identify the dualized directions, meaning we have to consider only subgroups of $\G$ and corresponding submatrices $E_{ab}, \wwt a, \wwt b$.

For general \pltp y the dilaton transformation rule was given in \cite{unge:pltp} and further studied in \cite{snohla:puzzle}. In the current notation we can write it as
\begin{align}
\wh\Phi(s, \hat x)=& \Phi(s, x)-\half\ln \Big| \det \left(\left(N +
\wh\Pi( \hat x) M\right)\wh a(\hat x)\right)\Big| \nonumber
\\&+\half \ln\Big| \det \left(\left({\bf 1} + \Pi( x) E(s)\right) a(x)\right)\Big|,
\label{dualdil}
\end{align}
$$
M=\mathcal{S}^T \cdot E(s)-\mathcal{Q}^T, \qquad N=\mathcal{P}^T-\mathcal{R}^T E(s).
$$
From the the two possible decompositions of elements of \dd\ 
$$
l=g(x)\wt h(\wt x)=\wh g(\hat x)\bar h(\bar x), \quad l\in\D, \quad
g\in \G, \, \wt h\in \wwt\G, \, \hat g\in \wh\G, \, \bar h\in \bG
$$
we can in principle express coordinates $x$ in terms of $\hat x$ and $\bar x$.
The expression is thus nonlocal in the sense that $\wh\Phi$ may depend also on \coor s $\bar x$ of $\bar{\G}$. For \plti ies we do not encounter this problem so it is plausible to use \eqref{dualdil} to calculate dilatons corresponding to B-shifts and $\beta$-shifts. For semi-Abelian \dd\ we find that dilaton does not change under B-shifts $I_B$, while under $I_\beta$ it transforms as
\begin{equation}\label{dualdil_beta}
\wh\Phi(s, x)= \Phi(s, x)-\half\ln \Big|\det {\bf 1}-\beta \cdot E_{ab}(s)\Big|.
\end{equation}
For duals $D_B=D_0 \cdot I_B$ and $D_\beta = D_0 \cdot I_\beta$ we get the correct dilaton by formula \eqref{dualdilaton} applied on the dilaton and \bkg\ obtained earlier from identities $I_B$ and $I_\beta$.

\section{Bianchi $V$ cosmology}\label{B_V}

As a warm up we shall study the well-known Bianchi $V$ cosmology. Let us
consider six-dimensional semi-Abelian \dd \footnote{{By ${\B}_V$, resp. ${\mathfrak b}_V$, we denote the Bianchi $V$ group,} resp. its Lie
algebra. {$\A$ and $\mathfrak a$ denote three dimensional
Abelian group and its Lie algebra respectively}. Similar notation will be used in the following sections.} $\D=({\B}_V |\A)$ whose Lie
algebra $\mathfrak{d}=\mathfrak b_V\bowtie \mathfrak a$ is spanned by
basis $(T_1,T_2,T_3,\wwt T^1,\wwt T^2,\wwt T^3)$. The non-trivial commutation relations of the generators of $\mathfrak b_V$ are
\begin{equation}
\label{comB5}
[ T_1, T_2]= T_2, \qquad [ T_1, T_3]= T_3 
.
\end{equation}
The group ${\B}_V$ is not semisimple and trace of its structure
constants does not vanish.

The \sm\ background\footnote{$E(s)$ is restored from $\cf(s,x)$ by setting group coordinates to zero.} is given by metric ($\mathcal{B}=0$ and $\cf = \mathcal{G}$)
\begin{equation} \label{F5_orig}
\cf (t,x_1) = \left(
\begin{array}{cccc}
 -1 & 0 & 0 & 0 \\
 0 & t^2 & 0 & 0 \\
 0 & 0 & e^{2 x_1} t^2 & 0 \\
 0 & 0 & 0 & e^{2 x_1} t^2 \\
\end{array}
\right).
\end{equation}
Left-invariant vector fields that satisfy \eqref{comB5} and generate symmetries of this background are
\begin{equation}
\nonumber
V_1=\partial_{x_1}-x_2\,\partial_{x_2}-x_3\,\partial_{x_3},\quad V_2=\partial_{x_2},\quad V_3=\partial_{x_3}.
\end{equation}
In fact, the \bkg\ is flat and torsionless so the standard \vbe\ are satisfied if we choose zero dilaton $\Phi = 0.$ This \bkg\ was studied already in \cite{GRV}, where it was first noticed that duals with respect to non-semisimple groups are not conformal. The related gravitational-gauge anomaly was later investigated in \cite{EGRSV}.

\subsection{\PL\ identities and dualities}
Mappings $C$ that preserve the algebraic structure of Manin triple {$(\cd,\mathfrak{b}_V,\ca)$} and generate \pl\ identities are given by matrices
$$
I_{1}=
\left(
\begin{array}{cccccc}
 1 & c_{12} & c_{13} & -c_{12} c_{15}-c_{13} c_{16} & c_{15} & c_{16} \\
 0 & c_{22} & c_{23} & -c_{15} c_{22}-c_{16} c_{23} & 0 & 0 \\
 0 & c_{32} & c_{33} & -c_{15} c_{32}-c_{16} c_{33} & 0 & 0 \\
 0 & 0 & 0 & 1 & 0 & 0 \\
 0 & 0 & 0 & \frac{c_{13} c_{32}-c_{12} c_{33}}{c_{22} c_{33}-c_{23} c_{32}} & \frac{c_{33}}{c_{22} c_{33}-c_{23} c_{32}} & \frac{c_{32}}{c_{23} c_{32}-c_{22} c_{33}} \\
 0 & 0 & 0 & \frac{c_{13} c_{22}-c_{12} c_{23}}{c_{23} c_{32}-c_{22} c_{33}} & \frac{c_{23}}{c_{23} c_{32}-c_{22} c_{33}} & \frac{c_{22}}{c_{22} c_{33}-c_{23} c_{32}} \\
\end{array}
\right),
$$
$$
I_{2}=
\left(
\begin{array}{cccccc}
 -1 & c_{12} & c_{13} & c_{12} c_{15}+c_{13} c_{16} & c_{15} & c_{16} \\
 0 & 0 & 0 & c_{12} c_{25}+c_{13} c_{26} & c_{25} & c_{26} \\
 0 & 0 & 0 & c_{12} c_{35}+c_{13} c_{36} & c_{35} & c_{36} \\
 0 & 0 & 0 & -1 & 0 & 0 \\
 0 & \frac{c_{36}}{c_{25} c_{36}-c_{26} c_{35}} & \frac{c_{35}}{c_{26} c_{35}-c_{25} c_{36}} & \frac{c_{16} c_{35}-c_{15} c_{36}}{c_{26} c_{35}-c_{25} c_{36}} & 0 & 0 \\
 0 & \frac{c_{26}}{c_{26} c_{35}-c_{25} c_{36}} & \frac{c_{25}}{c_{25} c_{36}-c_{26} c_{35}} & \frac{c_{16} c_{25}-c_{15} c_{26}}{c_{25} c_{36}-c_{26} c_{35}} & 0 & 0 \\
\end{array}
\right).
$$
One can see that {for $ c_{15}=c_{16}=0$ the matrix $I_1$ simplifies to the block form \eqref{IA} given by automorphisms of algebra $\mathfrak{b}_V$ that in general read
\begin{equation}\label{automorf 51}
A=\left(
\begin{array}{ccc}
 1 & a_{12} & a_{13} \\
 0 & a_{22} & a_{23} \\
 0 & a_{32} & a_{33} \\
\end{array}
\right).
\end{equation}
For $c_{12} = c_{13} = c_{23} = c_{32}=0$ and $c_{22}  = c_{33} = 1$ matrix $I_1$ reduces} to B-shift \eqref{IB} of the form
\begin{equation}\label{Bshift53}
I_B=
\left(
\begin{array}{cccccc}
 1 & 0 & 0 & 0 & c_{15} & c_{16} \\
 0 & 1 & 0 & -c_{15} & 0 & 0 \\
 0 & 0 & 1 & -c_{16} & 0 & 0 \\
 0 & 0 & 0 & 1 & 0 & 0 \\
 0 & 0 & 0 & 0 & 1 & 0 \\
 0 & 0 & 0 & 0 & 0 & 1 \\
\end{array}
\right).
\end{equation}
On the other hand, for $c_{12} = c_{13}= c_{26} = c_{35} = 0,$ $c_{25} =c_{36}=1$ matrix $I_2$ equals to
\begin{equation}\label{if1}
I_F=\left(
\begin{array}{cccccc}
 -1 & 0 & 0 & 0 & 0 & 0 \\
 0 & 0 & 0 & 0 &1 &0 \\
 0 & 0 & 0 & 0 &0 & 1 \\
 0 & 0 & 0 & -1 & 0 & 0 \\
 0 & 1 & 0 & 0 & 0 & 0 \\
 0 & 0 & 1 & 0 & 0 & 0 \\
\end{array}
\right).
\end{equation}
Matrix $I_F$ switches basis vectors $T_2, T_3$ and $\wwt T^2, \wwt T^3$. We identify its action as factorized duality with respect to Abelian subgroup generated by $T_2, T_3$. The change of sign of $T_1$ is necessary for being an automorphism of $\mathfrak b_V \bowtie \ca.$

To study models generated by $I_1$ and $I_2$ we decompose
these matrices into product of special elements of NATD group. Namely, we
note that $I_1$ can be written as
\begin{equation}\nonumber 
I_1 = I_A \cdot I_B
\end{equation}
where $I_A$ is given by  \eqref{automorf 51} and $I_B$ is the B-shift \eqref{Bshift53}. Similarly, $I_2$ can be decomposed as
\begin{equation}\nonumber
I_2 = I_{A_2} \cdot I_F \cdot I_{A_1}
\end{equation}
for automorphisms $A_1$ and $A_2$ of the form \eqref{automorf 51}. This decomposition is not unique. To identify relevant parameters of $I_2$ we choose the simplest possible $I_{A_1}$ while including the rest of the parameters in $I_{A_2}$ as follows:
\begin{equation}\label{automorf_51_A1_A2}
A_1=\left(
\begin{array}{ccc}
 1 & -c_{12} & -c_{13} \\
 0 & 1 & 0 \\
 0 & 0 & 1 \\
\end{array}
\right), \qquad
A_2=\left(
\begin{array}{ccc}
 1 & c_{15} & c_{16} \\
 0 & c_{25} & c_{26} \\
 0 & c_{35} & c_{36} \\
\end{array}
\right).
\end{equation}

Matrices generating \pltd ies can be obtained from those above by left-multiplication by matrix \eqref{Delta} representing canonical or ``full'' duality.
This way we get dual automorphisms
generated by
\begin{equation}\nonumber
D_A={D_0 \cdot I_A = }\begin{pmatrix} \mathbf{0}_d & (A^T)\-1\\ A & \mathbf{0}_d \end{pmatrix},
\end{equation}
dual B-shifts generated by
\begin{equation}\nonumber
D_B = D_0 \cdot I_B = \left(
\begin{array}{cccccc}
 0 & 0 & 0 & 1 & 0 & 0 \\
 0 & 0 & 0 & 0 & 1 & 0 \\
 0 & 0 & 0 & 0 & 0 & 1 \\
 1 & 0 & 0 & 0 & c_{15} & c_{16} \\
 0 & 1 & 0 & -c_{15} & 0 & 0 \\
 0 & 0 & 1 & -c_{16} & 0 & 0 \\
\end{array}
\right),
\end{equation} and factorized duality
\begin{equation}\label{5df}
D_F = D_0 \cdot I_F = \left(
\begin{array}{cccccc}
 0 & 0 & 0 & -1 & 0 & 0 \\
 0 & 1&0 & 0 & 0 & 0 \\
 0 &0&1 & 0 & 0 & 0 \\
 -1 & 0 & 0 & 0 & 0 & 0 \\
 0 & 0 & 0 & 0 &  1 & 0\\
 0 & 0 & 0 & 0 &  0 & 1 \\
\end{array}
\right)
\end{equation}
{that can be interpreted as Buscher duality with respect to $T_1$ accompanied by a change of sign in the dual coordinate.}

\subsection{Transformed \bkg s}

\subsubsection{B-shifts}

Let us now apply \plti ies on the \sm\ \eqref{F5_orig}. Plugging $I_1$ into formulas \eqref{pqrs}--\eqref{E0hat} we get rather complicated background tensor.
Nevertheless, $I_1$ decomposes as $I_1 = I_A \cdot I_B$ and we can get rid of the parameters that come from $I_A$ by a change of coordinates found by integrating the Jacobi matrix \eqref{Jac_transf}. Indeed, after coordinate transformation
\begin{align*}
y_1 = x_1, \quad
y_2 = -c_{12} e^{-x_1} + c_{22} x_2 + c_{32}x_3, \quad y_3 = -c_{13} e^{-x_1} + c_{23}x_2 + c_{33}x_3
\end{align*}
we find that the symmetric part of $\wh \cf$ equals to the original metric \eqref{F5_orig}. The antisymmetric part
\begin{equation}\nonumber
\wh{\mathcal{B}}(y_1)=\left(
\begin{array}{cccc}
 0 & 0 & 0 & 0 \\
 0 & 0 & -e^{y_1} c_{15} & -e^{y_1} c_{16} \\
 0 & e^{y_1} c_{15} & 0 & 0 \\
 0 & e^{y_1} c_{16} & 0 & 0 \\
\end{array}
\right)
\end{equation}
generated by the B-shift represents a torsionless B-field. Up to \coor {} \tfn s we would get the same \bkg\ using $I_B$ instead of the full $I_1$ so, from the point of view of \plti y, we consider these matrices equivalent. In other words, \plti y with respect to $I_B$ and $I_1$ is just a gauge transformation of the original \bkg, there is no change in the dilaton field, and $\wh\Phi=\Phi$ satisfies \vbe.

Background calculated using $D_1=D_0\cdot I_1$ is too extensive to be displayed. Nevertheless, a change of coordinates \eqref{Jac_transf} simplifies it to the form that one would obtain using $D_B$. Subsequent coordinate shift
\begin{align*}
\tilde x_1 & = c_{12} (\tilde{y}_{2}-c_{15})+c_{13} (\tilde{y}_{3}-c_{16})+\tilde{y}_{1}, \\
\tilde x_2 & = c_{22} (\tilde{y}_{2}-c_{15})+c_{23} (\tilde{y}_{3}-c_{16}),\\
\tilde x_3 & = c_{32} (\tilde{y}_{2}-c_{15})+c_{33} (\tilde{y}_{3}-c_{16})
\end{align*}
that agrees with the discussion in section \ref{factors} eliminates the parameters of $D_1$ completely, producing tensor
\begin{equation}\label{dmtz5DB}
\wwt\cf(t,\wt y_2,\wt y_3)=\left(
\begin{array}{cccc}
 -1 & 0 & 0 & 0 \\
 0 & \frac{t^2}{t^4+\tilde{y}_2^2+\tilde{y}_3^2} & \frac{\tilde{y}_2}{t^4+\tilde{y}_2^2+\tilde{y}_3^2} & \frac{\tilde{y}_3}{t^4+\tilde{y}_2^2+\tilde{y}_3^2} \\
 0 & -\frac{\tilde{y}_2}{t^4+\tilde{y}_2^2+\tilde{y}_3^2} & \frac{t^4+\tilde{y}_3^2}{t^2 \left(t^4+\tilde{y}_2^2+\tilde{y}_3^2\right)} & -\frac{\tilde{y}_2 \tilde{y}_3}{t^2 \left(t^4+\tilde{y}_2^2+\tilde{y}_3^2\right)} \\
 0 & -\frac{\tilde{y}_3}{t^4+\tilde{y}_2^2+\tilde{y}_3^2} & -\frac{\tilde{y}_2 \tilde{y}_3}{t^2 \left(t^4+\tilde{y}_2^2+\tilde{y}_3^2\right)} & \frac{t^4+\tilde{y}_2^2}{t^2 \left(t^4+\tilde{y}_2^2+\tilde{y}_3^2\right)} \\
\end{array}
\right).
\end{equation}
 The same \bkg\ can be obtained via full duality using $D_0$, and, as discussed in \cite{GRV,EGRSV}, it is not conformal. The standard \vbe\ cannot be satisfied by any dilaton $\wwt\Phi$. On the other hand, dilaton
$$\wwt\Phi(t,\wt y_2,\wt y_3)= -\frac{1}{2} \ln \left(t^2 \left(\wt y_2^2+\wt y_3^2+t^4\right)\right)$$
together with \bkg\ \eqref{dmtz5DB} satisfy Generalized Supergravity Equations \eqref{betaG}--\eqref{betaPhi} if we choose $\mathcal{J}=(0,2,0,0)$. Components of the Killing vector $\mathcal{J}$ are given by trace of structure constants\footnote{Compared to \cite{hokico}, in the present nomenclature the matrices $\cf$ representing background tensors are transposed. This results in change of sign of $\ci$.} of $\mathfrak{b}_{V}$ as $\mathcal{J}^a=f_{a i}^i$. The dilaton agrees with the formula \eqref{dualdilaton}, and we conclude that up to a \coor\ \tfn\ the \bkg\ \eqref{dmtz5DB} found using $D_B$ or $D_1$ is equivalent to non-Abelian T-dual investigated in \cite{hokico}.

\subsubsection{Factorized dualities}

{Using $I_2=I_{A_2}\cdot I_F \cdot I_{A_1}$ in formulas \eqref{pqrs}--\eqref{E0hat} we get background that can be brought to the form
\begin{equation}\nonumber 
\wh\cf (t,y_1)=\left(
\begin{array}{cccc}
 -1 & 0 & 0 & 0 \\
 0 & t^2 & -e^{y_1} c_{12} & -e^{y_1} c_{13} \\
 0 & e^{y_1} c_{12} & \frac{e^{2 y_1}}{t^2} & 0 \\
 0 & e^{y_1} c_{13} & 0 & \frac{e^{2 y_1}}{t^2} \\
\end{array}
\right)
\end{equation}
by coordinate transformation
\begin{align*}
y_1 = x_1, \quad
y_2 = -c_{15} e^{-x_1} + c_{25} x_2 + c_{35}x_3, \quad y_3 = -c_{16} e^{-x_1} + c_{26}x_2 + c_{36}x_3
\end{align*}
whose Jacobi matrix \eqref{Jac_transf} is determined by $A_2$ in \eqref{automorf_51_A1_A2}. The \bkg\ differs from $\wh\cf$ calculated using $I_F$ since $I_{A_1}$ changes $E(s)$ before the factorized duality is applied. However, the only difference is in the antisymmetric part $\wh{\mathcal{B}}$. For $I_2$ there is a torsionless B-field, while for $I_F$ the B-field vanishes completely.} The metric has vanishing scalar curvature but is not flat. Further \coor {} \tfn {}
\begin{align*}
t&=\sqrt{-2 u\, v+2 u+z_3^2+z_4^2}, & y_2&=u\, z_3,\\ y_1&=\frac{1}{2} \ln \left(\frac{-2 u\, v+2 u+z_3^2+z_4^2}{u^2}\right), &  y_3&=u\, z_4
\end{align*}
brings it to the Brinkmann form of plane parallel wave \cite{papa} with
$$
ds^2=2\frac{z_3^2+z_4^2}{u^2}du^2+2du\,dv+d z_3^2+d z_4^2.
$$

Corresponding dilaton
follows from the formula \eqref{dualdilaton} if the factorized
duality \eqref{if1} is interpreted as Buscher duality\footnote{{Followed by a change of sign in the spectator coordinate $x_1$.}} with respect
to two-dimensional Abelian subgroup generated by left-invariant
fields $V_2=\partial_{x_2},\ V_3=\partial_{x_3}$. Metric \eqref{F5_orig} is written in coordinates adapted to the action of this subgroup and for the duality given by $I_F$ we can write
$$\wh\Phi(t,x_1)=\frac{1}{2}\ln \det M=\frac{1}{2}\ln \det \left(
\begin{array}{cc}
\frac{e^{2 x_1}}{t^2} & 0 \\
  0 & \frac{e^{2 x_1}}{t^2} \\
\end{array}
\right)=-\ln \,t^2+2 y_1=-2 \ln u.$$
Dual dilaton for \bkg\ given by $I_2$ is derived from the altered {$E'(s)=\mathcal A_1 E(s) \mathcal A_1^T$} and differs from the previous expression by a constant. We again conclude that \bkg s found using $I_F$ and $I_2$ differ only by a coordinate and gauge transformation and can be considered equivalent. They satisfy \vbe, or Generalized
Supergravity Equations \eqref{betaG}--\eqref{betaPhi} where $\mathcal{J}$ is zero vector.

Background obtained by \PL {} transformation using matrix $D_F$
has the form
\begin{equation}\label{dualmtzIf5}
\wwt\cf(t,\wt y_2,\wt y_3)= \left(
\begin{array}{cccc}
 -1 & 0 & 0 & 0 \\
 0 & \frac{1}{t^2 \left(\wt y_2^2+\wt y_3^2+1\right)} & \frac{\wt y_2}{\wt y_2^2+\wt y_3^2+1} & \frac{\wt y_3}{\wt y_2^2+\wt y_3^2+1} \\
 0 & -\frac{\wt y_2}{\wt y_2^2+\wt y_3^2+1} & \frac{t^2 \left(\wt y_3^2+1\right)}{\wt y_2^2+\wt y_3^2+1} & -\frac{t^2 \wt y_2 \wt y_3}{\wt y_2^2+\wt y_3^2+1} \\
 0 & -\frac{\wt y_3}{\wt y_2^2+\wt y_3^2+1} & -\frac{t^2 \wt y_2 \wt y_3}{\wt y_2^2+\wt y_3^2+1} & \frac{t^2 \left(\wt y_2^2+1\right)}{\wt y_2^2+\wt y_3^2+1} \\
\end{array}
\right).
\end{equation}
The same \bkg\ is obtained using $D_2=D_0 \cdot I_2=D_0 \cdot I_{A_2}\cdot I_F \cdot I_{A_1}$ after change of coordinates
\begin{align*}
\tilde x_1 & = c_{15} (\tilde{y}_{2}-c_{12})+c_{16} (\tilde{y}_{3}-c_{13})+\tilde{y}_{1}, \\
\tilde x_2 & = c_{25} (\tilde{y}_{2}-c_{12})+c_{26} (\tilde{y}_{3}-c_{13}),\\
\tilde x_3 & = c_{35} (\tilde{y}_{2}-c_{12})+c_{36} (\tilde{y}_{3}-c_{13}).
\end{align*}
Thus, we are able to eliminate all parameters appearing in $D_2$. The \bkg\ is torsionless and together with dilaton
\begin{equation}\label{dualdil5DF}
\wwt\Phi(t,\wt y_2,\wt y_3) = -\half\ln \left(
t^2 \left(\wt y_2^2+\wt y_3^2+1\right) \right)
\end{equation}
satisfies \vbe, i.e. the Killing vector in the Generalized Supergravity Equations is zero. Explanation is that we can interpret the factorized duality
\eqref{5df} as Buscher duality of
\eqref{F5_orig}, this time with one-dimensional Abelian subgroup
generated by left-invariant field
$V_1=\partial_{x_1}-x_2\,\partial_{x_2}-x_3\,\partial_{x_3}$. In adapted \coor s $\{s_1,s_2,s_3,y_1\}$
\begin{equation}\nonumber 
 t=s_1, \quad x_1= y_1,\quad x_2=s_2 e^{-y_1},\quad x_3=s_3 e^{-y_1},
\end{equation}
where
$V_1=\partial_{y_1}$, the tensor \eqref{F5_orig} is manifestly invariant with respect to shifts in $y_1$ since
\begin{equation}\nonumber
\cf(s_1,s_2,s_3) = \left(
\begin{array}{cccc}
 -1 & 0 & 0 & 0 \\
 0 & s_1^2 & 0 & -s_1^2 s_2 \\
 0 & 0 & s_1^2 & -s_1^2 s_3 \\
 0 & -s_1^2 s_2 & -s_1^2 s_3 & s_1^2 \left(s_2^2+s_3^2+1\right) \\
\end{array}
\right).
\end{equation}
Buscher duality \wrt\ $y_1$ then restores the tensor \eqref{dualmtzIf5} and dilaton \eqref{dualdil5DF} agrees with formula \eqref{dualdilaton}.

To sum up, in this section we have shown that \bkg s emerging from general \plti ies or dualities differ from those obtained from special elements of NATD group only by a coordinate or gauge transformation. From now on we shall display results for these special elements and only comment on the general cases.

\section{Bianchi $III$ cosmology}\label{BIII}

Several results for Bianchi $III$ cosmology are similar to those for Bianchi $V$.
The algebra $\mathfrak{d}=\mathfrak b_{III} \bowtie\mathfrak a$ of six-dimensional semi-Abelian \dd\ {$({\B}_{III} |\A)$} is spanned by basis $(T_1,T_2,T_3,\wwt T^1,\wwt T^2,\wwt T^3)$. Non-trivial commutation relations of the generators of $\mathfrak b_{III}$ are
\begin{equation}
\label{comB3}
[ T_1, T_3]= -T_3,
\end{equation}
while $\mathfrak a$ is Abelian. The trace of structure constants does not vanish and group ${\B}_{III}$ is not semisimple. The background given by metric
\begin{equation}
\label{F3_orig}
\cf (t,x_1) = \left(
\begin{array}{cccc}
 -1 & 0 & 0 & 0 \\
 0 & t^2 & 0 & 0 \\
 0 & 0 & 1 & 0 \\
 0 & 0 & 0 & t^2 e^{- 2 x_1} \\
\end{array}
\right)
\end{equation}
is flat, torsionless, and invariant \wrt\ symmetries generated by left-invariant vector fields
\begin{equation}\nonumber 
V_1=\partial_{x_1}+x_3\,\partial_{x_3},\qquad
V_2=\partial_{x_2},\qquad V_3=\partial_{x_3}
\end{equation}
satisfying \eqref{comB3}. As the background is flat and torsionless the dilaton $\Phi$ 
can be chosen zero. Authors of \cite{GR} mention this \bkg\ in their analysis and note that its non-Abelian dual does not satisfy the standard \vbe.

\subsection{\PL\ identities and dualities}

\begin{table}
\begin{center}
\scriptsize
\begin{tabular}{|c||c|}
\hline
$({\B}_{III} |\A)$ & $C$ matrix  \\
\hline
$I_{1}$ & $
\left(
\begin{array}{cccccc}
 -1 & c_{12} & c_{13} & c_{14} & \frac{c_{14}-c_{13} c_{16}}{c_{12}} & c_{16} \\
 0 & 0 & 0 & \frac{c_{12}}{c_{52}} & \frac{1}{c_{52}} & 0 \\
 0 & 0 & 0 & c_{13} c_{36} & 0 & c_{36} \\
 0 & 0 & 0 & -1 & 0 & 0 \\
 0 & c_{52} & 0 & \frac{(c_{14}-c_{13} c_{16}) c_{52}}{c_{12}} & 0 & 0 \\
 0 & 0 & \frac{1}{c_{36}} & \frac{c_{16}}{c_{36}} & 0 & 0 \\
\end{array}
\right)$ \\
\hline
$I_{2}$ & $
\left(
\begin{array}{cccccc}
 -1 & c_{12} & c_{13} & c_{14} & \frac{c_{14}-c_{13} c_{16}}{c_{12}} & c_{16} \\
 0 & c_{22} & 0 & \frac{(c_{14}-c_{13} c_{16}) c_{22}}{c_{12}} & 0 & 0 \\
 0 & 0 & 0 & c_{13} c_{36} & 0 & c_{36} \\
 0 & 0 & 0 & -1 & 0 & 0 \\
 0 & 0 & 0 & \frac{c_{12}}{c_{22}} & \frac{1}{c_{22}} & 0 \\
 0 & 0 & \frac{1}{c_{36}} & \frac{c_{16}}{c_{36}} & 0 & 0 \\
\end{array}
\right)$ \\
\hline
$I_{3}$ & $
\left(
\begin{array}{cccccc}
 1 & c_{12} & c_{13} & c_{14} & -\frac{c_{14}+c_{13} c_{16}}{c_{12}} & c_{16} \\
 0 & 0 & 0 & -\frac{c_{12}}{c_{52}} & \frac{1}{c_{52}} & 0 \\
 0 & 0 & c_{33} & -c_{16} c_{33} & 0 & 0 \\
 0 & 0 & 0 & 1 & 0 & 0 \\
 0 & c_{52} & 0 & \frac{(c_{14}+c_{13} c_{16}) c_{52}}{c_{12}} & 0 & 0 \\
 0 & 0 & 0 & -\frac{c_{13}}{c_{33}} & 0 & \frac{1}{c_{33}} \\
\end{array}
\right)$ \\
\hline
$I_{4}$ & $
\left(
\begin{array}{cccccc}
 1 & c_{12} & c_{13} & c_{14} & -\frac{c_{14}+c_{13} c_{16}}{c_{12}} & c_{16} \\
 0 & c_{22} & 0 & \frac{(c_{14}+c_{13} c_{16}) c_{22}}{c_{12}} & 0 & 0 \\
 0 & 0 & c_{33} & -c_{16} c_{33} & 0 & 0 \\
 0 & 0 & 0 & 1 & 0 & 0 \\
 0 & 0 & 0 & -\frac{c_{12}}{c_{22}} & \frac{1}{c_{22}} & 0 \\
 0 & 0 & 0 & -\frac{c_{13}}{c_{33}} & 0 & \frac{1}{c_{33}} \\
\end{array}
\right)$ \\
\hline
$I_{5}$ & $
\left(
\begin{array}{cccccc}
 -1 & 0 & c_{13} & c_{13} c_{16} & c_{15} & c_{16} \\
 0 & 0 & 0 & 0 & \frac{1}{c_{52}} & 0 \\
 0 & 0 & 0 & c_{13} c_{36} & 0 & c_{36} \\
 0 & 0 & 0 & -1 & 0 & 0 \\
 0 & c_{52} & 0 & c_{15} c_{52} & 0 & 0 \\
 0 & 0 & \frac{1}{c_{36}} & \frac{c_{16}}{c_{36}} & 0 & 0 \\
\end{array}
\right)$ \\
\hline
$I_{6}$ & $
\left(
\begin{array}{cccccc}
 -1 & 0 & c_{13} & c_{13} c_{16} & c_{15} & c_{16} \\
 0 & c_{22} & 0 & c_{15} c_{22} & 0 & 0 \\
 0 & 0 & 0 & c_{13} c_{36} & 0 & c_{36} \\
 0 & 0 & 0 & -1 & 0 & 0 \\
 0 & 0 & 0 & 0 & \frac{1}{c_{22}} & 0 \\
 0 & 0 & \frac{1}{c_{36}} & \frac{c_{16}}{c_{36}} & 0 & 0 \\
\end{array}
\right)$ \\
\hline
$I_{7}$ & $
\left(
\begin{array}{cccccc}
 1 & 0 & c_{13} & -c_{13} c_{16} & c_{15} & c_{16} \\
 0 & 0 & 0 & 0 & \frac{1}{c_{52}} & 0 \\
 0 & 0 & c_{33} & -c_{16} c_{33} & 0 & 0 \\
 0 & 0 & 0 & 1 & 0 & 0 \\
 0 & c_{52} & 0 & -c_{15} c_{52} & 0 & 0 \\
 0 & 0 & 0 & -\frac{c_{13}}{c_{33}} & 0 & \frac{1}{c_{33}} \\
\end{array}
\right)$ \\
\hline
$I_{8}$ & $
\left(
\begin{array}{cccccc}
 1 & 0 & c_{13} & -c_{13} c_{16} & c_{15} & c_{16} \\
 0 & c_{22} & 0 & -c_{15} c_{22} & 0 & 0 \\
 0 & 0 & c_{33} & -c_{16} c_{33} & 0 & 0 \\
 0 & 0 & 0 & 1 & 0 & 0 \\
 0 & 0 & 0 & 0 & \frac{1}{c_{22}} & 0 \\
 0 & 0 & 0 & -\frac{c_{13}}{c_{33}} & 0 & \frac{1}{c_{33}} \\
\end{array}
\right)$ \\
\hline
\end{tabular}
\normalsize
\end{center}
\caption{PLT-identities of \dd\ $({\B}_{III} |\A)$.\label{tab:I_3}}
\end{table}

Tab.~\ref{tab:I_3} summarizes all eight types of solutions of \eqn s
\eqref{C_str_const} and \eqref{C_bilin} with structure constants
$F=\wh F$. These give rise to \PL\ identities and dualities of
$({\B}_{III} |\A)$. Nevertheless, all the identities are composed of
automorphisms \eqref{IA} with
\begin{equation}
\label{automorf 31}
A=\left(
\begin{array}{ccc}
 1 & a_{12} & a_{13} \\
 0 &  a_{22} & 0 \\ 0& 0 & a_{33} \\
\end{array}
\right),
\end{equation}
B-shifts of the form \eqref{Bshift53}, and factorized
dualities\footnote{$I_A$ and $I_B$ appear as special cases of $I_4$ and $I_8$, factorized dualities $I_{F_1},I_{F_2}$ and their composition appear in $I_1, I_2, I_3, I_5, I_6, I_7$ and their duals.}
\begin{equation}
\label{I3f1}
I_{F_1}=\left(
\begin{array}{cccccc}
 -1 & 0 & 0 & 0 & 0 & 0 \\
 0 & 1 & 0 & 0 & 0 & 0 \\
 0 & 0 & 0 & 0 & 0 & 1 \\
 0 & 0 & 0 & -1 & 0 & 0 \\
 0 & 0 & 0 & 0 & 1 & 0 \\
 0 & 0 & 1 & 0 & 0 & 0 \\
\end{array}
\right) 
\end{equation}
and
\begin{equation}\label{I3f2}
I_{F_2}=\left(
\begin{array}{cccccc}
 1 & 0 & 0 & 0 & 0 & 0 \\
 0 & 0 & 0 & 0 & 1 & 0 \\
 0 & 0 & 1 & 0 & 0 & 0 \\
 0 & 0 & 0 & 1 & 0 & 0 \\
 0 & 1 & 0 & 0 & 0 & 0 \\
 0 & 0 & 0 & 0 & 0 & 1 \\
\end{array}
\right)
.
\end{equation}
Matrices generating \pltd ies can be again obtained from those above
by left-multiplication  by the matrix \eqref{Delta} representing
full duality.

\subsection{Transformed \bkg s}

\subsubsection{B-shifts}

Using $I_B$ \eqref{Bshift53} in the formulas \eqref{pqrs}--\eqref{E0hat} we find that the background $\wh{\cf}$ has the same metric as the original model \eqref{F3_orig}. In addition to that, a torsionless $B$-field
\begin{equation}\label{F3_B_B}
\wh{\mathcal{B}}(x_1)=\left(
\begin{array}{cccc}
 0 & 0 & 0 & 0 \\
 0 & 0 & -c_{15} & - c_{16} e^{-x_1} \\
 0 & c_{15} & 0 & 0 \\
 0 & c_{16} e^{-x_1}  & 0 & 0 \\
\end{array}
\right)
\end{equation}
appears. This agrees with the interpretation of action of $I_B$ as gauge transformation. There is no change in the dilaton and $\wh{\Phi}=\Phi$.
With the full solutions $I_4$ and $I_8$ we get the same \bkg\ as for $I_B$. Indeed, both these matrices decompose as
$$I_4=I_A\cdot I_B, \qquad I_8=I_A \cdot I_B$$
with $I_A$ given by \eqref{automorf 31}. A linear change of coordinates \eqref{Jac_transf} thus restores the metric \eqref{F3_orig} and torsionless $B$-field \eqref{F3_B_B}\footnote{The parameter $c_{15}$ has to be replaced by $-\frac{c_{14}+c_{13} c_{16}}{c_{12}}$ for $I_4$.}.

Dual \bkg\ calculated using matrix $D_B=D_0\cdot I_B$ produces
tensor
\begin{equation}\label{F3_DB}
\wwt\cf(t,\wt x_3)=\left(
\begin{array}{cccc}
 -1 & 0 & 0 & 0 \\
 0 & \frac{t^2}{t^4+c_{15}^2 t^2+\left(c_{16}-\wt x_3\right){}^2} & \frac{t^2
   c_{15}}{t^4+c_{15}^2 t^2+\left(c_{16}-\wt x_3\right){}^2} &
   \frac{c_{16}-\wt x_3}{t^4+c_{15}^2 t^2+\left(c_{16}-\wt x_3\right){}^2} \\
 0 & -\frac{t^2 c_{15}}{t^4+c_{15}^2 t^2+\left(c_{16}-\wt x_3\right){}^2} &
   \frac{t^4+\left(c_{16}-\wt x_3\right){}^2}{t^4+c_{15}^2 t^2+\left(c_{16}-\wt x_3\right){}^2}
   & \frac{c_{15} \left(\wt x_3-c_{16}\right)}{t^4+c_{15}^2
   t^2+\left(c_{16}-\wt x_3\right){}^2} \\
 0 & \frac{\wt x_3-c_{16}}{t^4+c_{15}^2 t^2+\left(c_{16}-\wt x_3\right){}^2} & \frac{c_{15}
   \left(\wt x_3-c_{16}\right)}{t^4+c_{15}^2 t^2+\left(c_{16}-\wt x_3\right){}^2} &
   \frac{t^2+c_{15}^2}{t^4+c_{15}^2 t^2+\left(c_{16}-\wt x_3\right){}^2} \\
\end{array}
\right)
\end{equation}
whose curvature and torsion do not vanish. We can get rid of the parameter $c_{16}$ by shift in $\wt x_3$, but $c_{15}$ remains. As earlier, \bkg s calculated using $D_4=D_0 \cdot I_4= D_0 \cdot I_A \cdot I_B$ or $D_8=D_0 \cdot I_8= D_0 \cdot I_A \cdot I_B$ differ from $\wwt{\cf}$ only by a transformation of coordinates. {For nonzero $c_{15}$ the tensor $\wwt \cf$ is not the same as non-Abelian dual of \eqref{F3_orig} that can be found using $D_0$. Nevertheless, if we understand the duality \wrt\ $D_B=D_0 \cdot I_B$ as full duality applied to \bkg\ changed by $I_B$, the correct dilaton can be found from \eqref{dualdilaton} as
\begin{equation}\label{Phi3_DB}
\wwt\Phi(t,\wt x_3)=- \frac{1}{2}\ln\left(t^4+c_{15}^2t^2+\left(c_{16}-\wt x_3\right)^2\right).
\end{equation}
Such $\wwt\Phi$ satisfies the Generalized Supergravity Equations for Killing vector $\mathcal{J}=(0,-1,0,0)$ whose components are given by trace of structure constants of $\mathfrak{b}_{III}$ as suggested in \cite{hokico}.}

\subsubsection{Factorized dualities}

\PL\ identities \eqref{I3f1} and \eqref{I3f2} can be interpreted as Buscher dualities with respect to one-dimensional Abelian subgroups generated by left-invariant fields
$V_3=\partial_{x_3}$ resp. $V_2=\partial_{x_2}$.

Dualization \wrt\ $V_2$ does not change $\cf$ at all due to the form of the metric \eqref{F3_orig}. The background is invariant \wrt\ $I_{F_2}$. Its dual given by $D_{F_2}=D_0 \cdot I_{F_2}$ needs to be understood as dual \wrt\ non-Abelian group generated by $V_1, V_3$ that is not semisimple. 
The \bkg\ and dilaton are the same as for the full duality $D_0$. We can read them from \eqref{F3_DB}, \eqref{Phi3_DB} setting $c_{15}=c_{16}=0$. The same results, up to a coordinate or gauge transformation, are obtained for the full solutions $I_3$, $I_7$, see Tab. \ref{tab:I_3}, and their duals $D_3$, $D_7$ since
$$
I_3=I_{A_2}\cdot I_{F_2} \cdot I_B \cdot I_{A_1}, \qquad I_7 = I_{A_2} \cdot I_{F_2} \cdot I_B.
$$

{Dualization \wrt\ $V_3$, i.e. \plti y $I_{F_1}$,} produces metric
\begin{equation}\label{F3_IF1}
\wh\cf (t,x_1)=\left(
\begin{array}{cccc}
 -1 & 0 & 0 & 0 \\
 0 & t^2 & 0 & 0 \\
 0 & 0 & 1 & 0 \\
 0 & 0 & 0 & \frac{e^{-2 x_1}}{t^2} \\
\end{array}
\right)
\end{equation}
{whose scalar curvature vanishes.} In coordinates
\begin{align*}
t&=\sqrt{z_3^2-2 u (v-1)}, & x_2&=z_4,\\ x_1&=-\frac{1}{2} \ln \left(\frac{z_3^2-2 u (v-1)}{u^2}\right), & x_3&=u z_3
\end{align*}
it acquires the Brinkmann form of a plane parallel wave with
$$ ds^2=2\frac{z_3^2}{u^2}du^2+2du\,dv+d z_3^2+d z_4^2.$$
As expected, dilaton calculated via formula \eqref{dualdilaton}
$$ \wh\Phi(t,x_1)=\frac{1}{2}\ln \det M= \frac{1}{2}\ln \det \left(
\begin{array}{cc}
1 & 0 \\
  0 & \frac{e^{-2 x_1}}{t^2} \\
\end{array}
\right)=  -\half \ln\ t^2 - x_1
$$
satisfies \vbe, or Generalized Supergravity Equations
with $\mathcal{J}=0$, since we have dualized \wrt\ Abelian subgroup of $\B_{III}$.
\plti ies $I_1$, $I_2$, $I_5$ and $I_6$ decompose as
\begin{align*}
I_1 & =I_{A_2}\cdot I_{F_1} \cdot I_{F_2} \cdot I_{A_1}, & I_2 & = I_{A_2} \cdot I_{F_1}\cdot I_{B} \cdot I_{A_1},\\
I_5 & =I_{A_2}\cdot I_{F_1} \cdot I_{F_2} \cdot I_{A_1}, & I_6 & =I_{A_2} \cdot I_{F_1} \cdot I_{B} \cdot I_{A_1}.
\end{align*}
Resulting \bkg s differ from \eqref{F3_IF1} only by a change of coordinates and torsionless B-field of the form \eqref{F3_B_B} and can be found in Tab. \ref{tab:app_B3}.

Dual \bkg\ produced by $D_{F_1}=D_0\cdot I_{F_1}$ reads
\begin{equation}\nonumber 
\wwt\cf(t,\wt x_3)=\left(
\begin{array}{cccc}
 -1 & 0 & 0 & 0 \\
 0 & \frac{1}{t^2 \left(\wt x_3^2+1\right)} & 0 & -\frac{\wt x_3}{\wt x_3^2+1} \\
 0 & 0 & 1 & 0 \\
 0 & \frac{\wt x_3}{\wt x_3^2+1} & 0 & \frac{t^2}{\wt x_3^2+1} \\
\end{array}
\right).
\end{equation}
Together with the dilaton
\begin{equation}\nonumber 
\wwt\Phi(t,\wt x_3)=-\frac{1}{2}\ln\left(t^2\left(\wt x_3^2+1\right)\right)
\end{equation}
found from \eqref{dualdilaton} this \bkg\ satisfies \vbe. Factorized duality
given by $D_{F_1}$ can be once again interpreted as Buscher
duality with respect to symmetry generated by $V_1,V_2$. The same result is obtained for $D_5=D_0 \cdot I_5$. For $D_1$, $D_2$, $D_6$ the tensor $\wwt \cf$ and dilaton $\wwt\Phi$ contain a parameter that cannot be eliminated by coordinate or gauge transformation. Interested reader may find its full form in Tab. \ref{tab:app_B3} in the Appendix.

\section{Bianchi $VI_\kappa$ cosmology}

Semi-Abelian \dd\ $\D=({\B}_{VI_\kappa} |\A)$ has Lie algebra {$\mathfrak{d}=\mathfrak b_{VI_\kappa}\bowtie\mathfrak a$} spanned by basis $(T_1,T_2,T_3,\wwt T^1,\wwt T^2,\wwt T^3)$ and the nontrivial \comrel s of $\mathfrak b_{VI_\kappa}$ are\footnote{Note that for $\kappa = 0$, or $\kappa = 1$, these are \comrel s of $\mathfrak b_{III}$, or $\mathfrak b_{V}$, respectively. The case $\kappa = -1$ will be treated separately in section \ref{B_VI-1}.}
\begin{equation}
\label{comB6} [ T_1, T_2]=\kappa\, T_2, \qquad [T_1, T_3]= T_3,\qquad
\kappa\,\neq\, - 1.
\end{equation}
Trace of structure constants does not vanish and group
{${\B}_{VI_\kappa}$} is not semisimple. In the parametrization used
in \cite{hokico} Bianchi $VI_\kappa$ cosmology is given by metric
\begin{equation}\label{mtz61}
\cf(t,x_1)=\left(
\begin{array}{cccc}
 -e^{-4 \Phi (t)}a_1(t)^2{a_2}(t)^2{a_3}(t)^2 & 0 & 0 & 0 \\
 0 &a_1(t)^2 & 0 & 0 \\
 0 & 0 & e^{2 \kappa x_1}{a_2}(t)^2 & 0 \\
 0 & 0 & 0 & e^{2 x_1}{a_3}(t)^2 \\
\end{array}
\right)
\end{equation}
where the functions $a_i(t)$ are
\begin{align}\label{B6k_ai}
{a_1}(t)&=e^{\Phi(t)}\left(\frac{{p_1}}{\kappa +1}\right)^{\frac{\kappa^2+1}{(\kappa +1)^2}}e^{\frac{(\kappa -1) {p_2} t}{2 (\kappa +1)}}\sinh ^{-\frac{\kappa ^2+1}{(\kappa+1)^2}}({p_1} t),\nonumber\\
{a_2}(t)&= e^{\Phi(t)}\left(\frac{{p_1}}{\kappa +1}\right)^{\frac{\kappa}{\kappa +1}} e^{\frac{{p_2} t}{2}} \sinh ^{-\frac{\kappa}{\kappa +1}}({p_1} t),\\
{a_3}(t)&= e^{\Phi(t)}\left(\frac{{p_1}}{\kappa +1}\right)^{\frac{1}{\kappa +1}}e^{\frac{-{p_2} t}{2}} \sinh ^{-\frac{1}{\kappa +1}}({p_1} t).\nonumber
\end{align}
The background is invariant with respect to symmetry generated by left-invariant vector fields
\begin{equation} \nonumber 
V_1=\partial_{x_1}-\kappa\,x_2\,\partial_{x_2}-x_3\partial_{x_3}, \qquad V_2=\partial_{x_2}, \qquad V_3=\partial_{x_3}
\end{equation}
satisfying \eqref{comB6}. For dilaton $\Phi (t)=c_1 t$ the \vbe\
reduce to condition
\begin{equation}\nonumber
c_1^2=\frac{ \left(\kappa^2+\kappa +1\right) p_1^2}{(\kappa +1)^2}-\frac{p_2^2}{4}.
\end{equation}
The background is torsionless and for $c_1=0$ also Ricci flat.

\subsection{\PL\ identities and dualities}
\PL\ identities of \dd\ $({\B}_{VI_\kappa} |\A)$ are given by
matrices
\begin{equation}\nonumber 
I_1=
\left(\begin{array}{cccccc}
 1 & c_{12} & c_{13} & -c_{12} c_{15}-c_{13} c_{16} & c_{15} & c_{16} \\
 0 & c_{22} & 0 & -c_{15} c_{22} & 0 & 0 \\
 0 & 0 & c_{33} & -c_{16} c_{33} & 0 & 0 \\
 0 & 0 & 0 & 1 & 0 & 0 \\
 0 & 0 & 0 & -\frac{c_{12}}{c_{22}} & \frac{1}{c_{22}} & 0 \\
 0 & 0 & 0 & -\frac{c_{13}}{c_{33}} & 0 & \frac{1}{c_{33}} \\
\end{array}
\right)
\end{equation}
\begin{equation}\nonumber 
I_{2}= \left(
\begin{array}{cccccc}
 -1 & c_{12} & c_{13} & c_{12} c_{15}+c_{13} c_{16} & c_{15} & c_{16} \\
 0 & 0 & 0 & c_{12} c_{25} & c_{25} & 0 \\
 0 & 0 & 0 & c_{13} c_{36} & 0 & c_{36} \\
 0 & 0 & 0 & -1 & 0 & 0 \\
 0 & \frac{1}{c_{25}} & 0 & \frac{c_{15}}{c_{25}} & 0 & 0 \\
 0 & 0 & \frac{1}{c_{36}} & \frac{c_{16}}{c_{36}} & 0 & 0 \\
\end{array}
\right).\end{equation} The algebra $\mathfrak{b}_{VI_\kappa}$ admits
automorphisms \eqref{automorf 31} and matrices $I_A$ of the form
\eqref{IA} are among the special cases of $I_1$. Clearly, $I_1$ is a
product $I_1=I_A \cdot I_B$ of automorphisms and B-shifts
\eqref{Bshift53}. Matrix $I_2$ can be written as $I_2=I_{A_2} \cdot
I_F \cdot I_{A_1}$ where $I_F$ is the factorized duality \eqref{if1}
and $I_{A_1}, I_{A_2}$ are given by automorphisms
\eqref{automorf_51_A1_A2}. \pltd ies are obtained by multiplication
by $D_0$.

\subsection{Transformed \bkg s}

\subsubsection{B-shifts}

{Using $I_1$ directly in formulas \eqref{pqrs}--\eqref{E0hat} we get
rather complicated background tensor. However, since $I_1$ splits as
$I_1=I_A\cdot I_B$, the dependence of $\wh\cf$ on the parameters
appearing in $I_A$ can be eliminated by transformation
\eqref{Jac_transf}. The \bkg\ obtained using $I_1$ is equivalent to 
that obtained by B-shift \eqref{Bshift53} and reads}
\begin{equation}\nonumber 
\wh\cf(t,x_1)=\left(
\begin{array}{cccc}
 -e^{-4 \Phi(t)} a_1(t)^2 a_2(t)^2 a_3(t)^2 & 0 & 0 & 0 \\
 0 & a_1(t)^2 & -e^{\kappa x_{1}} c_{15} & -e^{x_{1}} c_{16} \\
 0 & e^{\kappa x_{1}} c_{15} & e^{2 \kappa x_{1}} a_2(t)^2 & 0 \\
 0 & e^{x_{1}} c_{16} & 0 & e^{2 x_{1}} a_3(t)^2 \\
\end{array}
\right).
\end{equation}
Beside the original metric \eqref{mtz61} we have obtained a torsionless B-field. Together with the original dilaton $\Phi (t_{1})=c_1 t$ the \bkg\ satisfies \vbe.

{Dual background $\wwt\cf$ calculated using $D_1=D_0\cdot I_1=D_0
\cdot I_A \cdot I_B$ is again too complicated to display.
Nevertheless, linear transformation of coordinates
\eqref{Jac_transf} followed by shift in $\tilde{y}_2, \tilde{y}_3$
simplifies the background to}
\begin{align*}
&\wwt{\cf}(t,\wt y_2,\wt y_3)=\\
&\left(
\begin{array}{cccc}
 -e^{-4 \Phi (t)} a_1(t)^2 a_2(t)^2 a_3(t)^2 & 0 & 0 & 0 \\
 0 & \frac{a_2(t)^2 a_3(t)^2}{\Delta} & \frac{\kappa a_3(t)^2 \tilde{y}_{2}}{\Delta} & \frac{a_2(t)^2 \tilde{y}_{3}}{\Delta} \\
 0 & -\frac{\kappa a_3(t)^2 \tilde{y}_{2}}{\Delta} & \frac{a_1(t)^2 a_3(t)^2+\tilde{y}_{3}^2}{\Delta} & -\frac{\kappa \tilde{y}_{2} \tilde{y}_{3}}{\Delta} \\
 0 & -\frac{a_2(t)^2 \tilde{y}_{3}}{\Delta} & -\frac{\kappa \tilde{y}_{2} \tilde{y}_{3}}{\Delta} & \frac{a_1(t)^2 a_2(t)^2+\kappa^2 \tilde{y}_{2}^2}{\Delta} \\
\end{array}
\right),
\end{align*}
where
$$
\Delta = a_1(t)^2 a_2(t)^2 a_3(t)^2+\kappa^2 \tilde{y}_{2}^2 a_3(t)^2+a_2(t)^2 \tilde{y}_{3}^2.
$$
These results are the same as results obtained by full duality
$D_0$. Dual dilaton
$$
\wwt\Phi(t,\wt y_2, \wt y_3) = c_1 t - \frac{1}{2} \ln \Delta
$$
found from formula \eqref{dualdilaton} satisfies the generalized
supergravity equations \eqref{betaG}--\eqref{betaPhi} where
components of Killing vector $\mathcal{J}=(0,\kappa+1,0,0)$ correspond to
trace of structure constants of $\mathfrak{b}_{VI_\kappa}$.
Dualization with respect to $D_1$ can be treated as canonical
duality in spite of the fact that it contains also B-shifts and
automorphisms.

\subsubsection{Factorized dualities}

\PL\ identity $I_F$ in \eqref{if1} can be interpreted as Buscher
duality with respect to two-dimensional Abelian subgroup generated
by left-invariant fields $V_2=\partial_{x_2},\ V_3=\partial_{x_3}$.
Resulting curved background
\begin{equation}\label{mtzif6}
\wh\cf(t,x_1)=\left(
\begin{array}{cccc}
 -e^{-4 \Phi(t)}a_1(t)^2{a_2}(t)^2{a_3}(t)^2 & 0 & 0 & 0 \\
 0 &a_1(t)^2 & 0 & 0 \\
 0 & 0 & \frac{e^{2 \kappa  x_1}}{{a_2}(t)^2} & 0 \\
 0 & 0 & 0 & \frac{e^{2 x_1}}{{a_3}(t)^2} \\
\end{array}
\right)
\end{equation}
and dilaton
\begin{equation}\label{dual_dil_61_1}
\wh\Phi(t,x_1)=c_1 t+\frac{1}{2}\ln \left(\frac{e^{2 (\kappa+1)  x_1}}{{a_2}(t)^2 {a_3}(t)^2}\right)
\end{equation}
calculated by formula \eqref{dualdilaton} satisfy \vbe \  
 with vanishing Killing vector $\mathcal{J}$ since we have dualized with respect to Abelian group.

{\plti y $I_2=I_{A_2} \cdot I_F \cdot I_{A_1}$ gives a background
whose metric can be brought to the form \eqref{mtzif6} by coordinate
transformation \eqref{Jac_transf}. There is also a torsionless
$B$-field depending on constants coming from $I_{A_1}$ that
transforms $E(s)$ to $E'(s)=\mathcal A_1 E(s) \mathcal A_1^T$.
Dilaton found by \eqref{dualdilaton} from $E'(s)$ differs from
\eqref{dual_dil_61_1} only by a constant shift and we may conclude
that results of duality with respect to $I_2$ deviate from those
obtained by $I_F$ only by coordinate and gauge transformation.}

After a suitable  coordinate transformation we find
that both matrices $D_F=D_0\cdot I_F$ and $D_2=D_0\cdot I_{A_2}
\cdot I_F \cdot I_{A_1}$ produce background
\begin{equation}\nonumber 
\wwt\cf(t,\wt y_2, \wt y_3)=\left(
\begin{array}{cccc}
 -e^{-4 \Phi (t)} a_1^2 a_2^2 a_3^2 & 0 & 0 & 0 \\
 0 & \frac{1}{\Delta} & \frac{\kappa a_2^2 \wt y_{2}}{\Delta} & \frac{a_3^2 \wt y_{3}}{\Delta} \\
 0 & -\frac{\kappa a_2^2 \wt y_{2}}{\Delta} & \frac{a_2^2 \left(a_1^2+a_3^2 \wt y_{3}^2\right)}{\Delta} & -\frac{\kappa a_2^2 a_3^2 \wt y_{2} \wt y_{3}}{\Delta} \\
 0 & -\frac{a_3^2 \wt y_{3}}{\Delta} & -\frac{\kappa a_2^2 a_3^2 \wt y_{2} \wt y_{3}}{\Delta} & \frac{a_3^2 \left(a_1^2+\kappa^2 a_2^2 \wt y_{2}^2\right)}{\Delta} \\
\end{array}
\right)
\end{equation}
where
$$\Delta = {a_1}(t)^2+\kappa^2 {a_2}(t)^2 \wt y_{2}^2+{a_3}(t)^2 \wt y_{3}^2.$$
This \bkg\ is the same as the one that would be obtained by performing Buscher duality with respect to symmetry generated by $V_1$. Dilaton
\begin{equation}\nonumber 
\wwt\Phi(t,\wt y_2, \wt y_3)=c_1 t - \frac{1}{2} \ln \Delta
\end{equation}
satisfies ordinary \vbe.

Let us note that results of this section hold also for $\kappa=0,1$,
i.e. for Bianchi $III$ and Bianchi $V$.
Dualities with respect to these groups were treated in sections
\ref{B_V} and \ref{BIII} with different initial backgrounds.

\section{Bianchi $VI_{-1}$ cosmology}\label{B_VI-1}

For Bianchi $VI_{-1}$ cosmology we shall consider Manin triple
$(\mathfrak d,\mathfrak b_{VI_{-1}},\ca)$ 
whose algebraic structure is given by \comrel s \eqref{comB6} with
$\kappa=-1$. Structure coefficients of Lie algebra $\mathfrak
b_{VI_{-1}}$ are traceless and the group $\B_{VI_{-1}}$ is not semisimple. Metric
has the form \eqref{mtz61} with functions
\begin{equation}\label{B6-1_ai}
a_1(t) = \sqrt{p_1} \exp \left( \left(\frac{e ^{2 p_2 t}+p_1 t}{2}\right) + \Phi (t) \right), \quad a_2(t)= a_3(t) = \sqrt{p_2} e^{\frac{p_2 t}{2}+\Phi(t)}
\end{equation}
and dilaton is again $\Phi (t)=c_1 t$.
The \vbe\ are satisfied if
$$
c_1^2=\frac{1}{4}(2 p_1 p_2 + p_2^2).
$$

\subsection{\PL\ identities and dualities}

\PL\ identities of \dd\ $({\B}_{VI_{-1}} |\A)$ are given by matrices
\begin{equation*}
I_{1}=
\left(
\begin{array}{cccccc}
 -1 & c_{12} & c_{13} & c_{12} c_{15}+c_{13} c_{16} & c_{15} & c_{16} \\
 0 & 0 & \frac{c_{32}}{c_{32} c_{56}-c_{36} c_{52}} & \frac{c_{16} c_{32}-c_{12} c_{36}}{c_{32} c_{56}-c_{36} c_{52}} & \frac{c_{36}}{c_{36} c_{52}-c_{32} c_{56}} & 0 \\
 0 & c_{32} & 0 & c_{15} c_{32}+c_{13} c_{36} & 0 & c_{36} \\
 0 & 0 & 0 & -1 & 0 & 0 \\
 0 & c_{52} & 0 & c_{15} c_{52}+c_{13} c_{56} & 0 & c_{56} \\
 0 & 0 & \frac{c_{52}}{c_{36} c_{52}-c_{32} c_{56}} & \frac{c_{16} c_{52}-c_{12} c_{56}}{c_{36} c_{52}-c_{32} c_{56}} & \frac{c_{56}}{c_{32} c_{56}-c_{36} c_{52}} & 0 \\
\end{array}
\right)
\end{equation*}
and
\begin{equation*}
I_{2}= \left(
\begin{array}{cccccc}
 1 & c_{12} & c_{13} & -c_{12} c_{15}-c_{13} c_{16} & c_{15} & c_{16} \\
 0 & \frac{c_{33}}{c_{33} c_{55}-c_{35} c_{53}} & 0 & \frac{c_{15} c_{33}-c_{13} c_{35}}{c_{35} c_{53}-c_{33} c_{55}} & 0 & \frac{c_{35}}{c_{35} c_{53}-c_{33} c_{55}} \\
 0 & 0 & c_{33} & -c_{16} c_{33}-c_{12} c_{35} & c_{35} & 0 \\
 0 & 0 & 0 & 1 & 0 & 0 \\
 0 & 0 & c_{53} & -c_{16} c_{53}-c_{12} c_{55} & c_{55} & 0 \\
 0 & \frac{c_{53}}{c_{35} c_{53}-c_{33} c_{55}} & 0 & \frac{c_{15} c_{53}-c_{13} c_{55}}{c_{33} c_{55}-c_{35} c_{53}} & 0 & \frac{c_{55}}{c_{33} c_{55}-c_{35} c_{53}} \\
\end{array}
\right).
\end{equation*}
{As special cases we find two types of automorphisms $I_A$ and $I_{A'}$ given by
\begin{equation}\nonumber 
A=\left(
\begin{array}{ccc}
 1 & a_{12} & a_{13}\\
 0 & {a_{22}} & 0 \\
 0 & 0 & {a_{33}} \\
\end{array}
\right), \qquad A'=\left(
\begin{array}{ccc}
-1 & a_{12} & a_{13}\\
 0 & 0 & {a_{23}} \\
 0 & {a_{32}} & 0 \\
\end{array}
\right),
\end{equation}
B-shifts generated by matrix
\begin{equation}\label{IB2}
I_B=\left(
\begin{array}{cccccc}
 1 & 0 & 0 & 0 & {b_{12}} & {b_{13}} \\
 0 & 1 & 0 & -{b_{12}} & 0 & {b_{23}} \\
 0 & 0 & 1 & -{b_{13}} & -{b_{23}} & 0 \\
 0 & 0 & 0 & 1 & 0 & 0 \\
 0 & 0 & 0 & 0 & 1 & 0 \\
 0 & 0 & 0 & 0 & 0 & 1 \\
\end{array}
\right),
\end{equation}
$\beta$-shifts
\begin{equation}\nonumber 
I_\beta=\left(
\begin{array}{cccccc}
 1 & 0 & 0 & 0 & 0 & 0 \\
 0 & 1 & 0 & 0 & 0 & 0 \\
 0 & 0 & 1 & 0 & 0 & 0 \\
 0 & 0 & 0 & 1 & 0 & 0 \\
0 &0 & {\beta_{23}} &  0 & 1 & 0 \\
0 & -{\beta_{23}} & 0& 0 & 0 & 1 \\
\end{array}
\right),
\end{equation}
and factorized dualities \eqref{if1}. To analyze results following from application of \plti ies $I_1$ and $I_2$ it is helpful to find their decomposition into products of special elements of NATD group. Depending on the values of parameters the matrices can be written as
\begin{equation}\nonumber
I_1  =
\left\{
\begin{array}{ll}
      I_{A'} \cdot I_B \cdot I_\beta & \text{ for }c_{56}\neq 0 \\
      I_{A'} \cdot I_\beta \cdot I_B & \text{ for }c_{56}= 0, c_{32}\neq 0 \\
      I_A\cdot I_B \cdot I_F & \text{ for }c_{56}= c_{32}=0
\end{array}
\right.
\end{equation}
and
\begin{equation}\nonumber
I_2  =
\left\{
\begin{array}{ll}
      I_{A} \cdot I_B \cdot I_\beta & \text{ for }c_{55}\neq 0 \\
      I_{A} \cdot I_\beta \cdot I_B & \text{ for }c_{55}= 0, c_{33}\neq 0 \\
      I_{A'}\cdot I_B \cdot I_F & \text{ for }c_{55}= c_{33}=0
\end{array}
\right.
\end{equation}
for some $I_A$, $I_{A'}$, $I_B$, $I_\beta$ and $I_F$. The parameters rising from $I_A$ and $I_{A'}$ can be again eliminated by coordinate transformation \eqref{Jac_transf}. It is thus sufficient to discuss \bkg s obtained from $I_B$, $I_\beta$, $I_F$ and their products. Multiplying these matrices by $D_0$ we get \pltd ies.

\subsection{Transformed \bkg s}

\subsubsection{B-shifts}

Transformed \bkg
\begin{equation}\label{B6-1_B}
\wh\cf(t,x_1)=\left(
\begin{array}{cccc}
 -e^{-4 \Phi(t)}a_1(t)^2{a_2}(t)^4 & 0 & 0 & 0 \\
 0 &a_1(t)^2 & -{b_{12}} e^{-x_1} & -{b_{13}} e^{x_1} \\
 0 & {b_{12}} e^{-x_1} & e^{-2 x_1}{a_2}(t)^2 & -{b_{23}} \\
 0 & {b_{13}} e^{x_1} & {b_{23}} & e^{2 x_1}{a_2}(t)^2 \\
\end{array}
\right)
\end{equation}
given by  B-shift differs from original $\cf$ by a torsionless B-field and  together with the dilaton $\wh\Phi=c_1 t$ satisfies \vbe.

Coordinate shifts eliminate ${b_{12}},{b_{13}}$ in the dual obtained from $D_0\cdot I_B$ so it reads
\begin{equation}\nonumber 
\wwt\cf(t, \wt y_2, \wt y_3)=\left(
\begin{array}{cccc}
 -e^{-4 \Phi (t)} a_1^2 a_2^4 & 0 & 0 & 0 \\
 0 & \frac{a_2^4+b_{23}^2}{\Delta} & -\frac{\wt y_2 a_2^2+b_{23} \wt y_3}{\Delta} & \frac{a_2^2 \wt y_3-b_{23} \wt y_2}{\Delta} \\
 0 & \frac{a_2^2 \wt y_2-b_{23} \wt y_3}{\Delta} & \frac{a_1^2 a_2^2+\wt y_3^2}{\Delta} & \frac{b_{23} a_1^2+\wt y_2 \wt y_3}{\Delta} \\
 0 & -\frac{\wt y_3 a_2^2+b_{23} \wt y_2}{\Delta} & \frac{\wt y_2 \wt y_3-b_{23} a_1^2}{\Delta} & \frac{a_1^2 a_2^2+\wt y_2^2}{\Delta} \\
\end{array}
\right)
\end{equation}
where
$$
\Delta=\left(a_2(t)^4+b_{23}^2\right) a_1(t)^2+a_2(t)^2 \left(\wt y_2^2+\wt y_3^2\right).
$$
The constant ${b_{23}}$ remains. For dilaton
$$\wwt\Phi(t, \wt y_2, \wt y_3)=c_1  t-\frac{1}{2} \ln \Delta$$
\vbe\ are satisfied. Vanishing of vector $\mathcal{J}$ corresponds to the fact that structure constants of $\mathfrak{b}_{VI_{-1}}$ are traceless.

\subsubsection{$\beta$-shifts}

Background given by  $\beta$-shift is
\begin{equation}\label{B6-1_Beta}
\wh\cf(t,x_1)=\left(
\begin{array}{cccc}
 -e^{-4 \Phi(t)}a_1(t)^2{a_2}(t)^4 & 0 & 0 & 0 \\
 0 &a_1(t)^2 & 0 & 0 \\
 0 & 0 & \frac{e^{-2 x_1}{a_2}(t)^2}{\beta_{23}^2{a_2}(t)^4+1} & \frac{{\beta_{23}}
  {a_2}(t)^4}{\beta_{23}^2{a_2}(t)^4+1} \\
 0 & 0 & -\frac{{\beta_{23}}{a_2}(t)^4}{\beta_{23}^2{a_2}(t)^4+1} & \frac{e^{2 x_1}
  {a_2}(t)^2}{\beta_{23}^2{a_2}(t)^4+1} \\
\end{array}
\right)
\end{equation}
and together with the dilaton calculated by formula \eqref{dualdil_beta}
$$ \wh\Phi(t)=c_1 t - \frac{1}{2} \ln \left(\beta_{23}^2 a_2(t)^4+1 \right)
$$
satisfy \vbe. Although matrices $I_B$ and $I_\beta$ do not commute, \bkg s obtained from $I_B \cdot I_\beta$ and $I_\beta \cdot I_B$ are the same and differ from $\wh \cf$ in \eqref{B6-1_Beta} only by a torsionless B-field.

The dual obtained from $D_0 \cdot I_\beta$ is
\begin{align*}
&\wwt\cf(t,\wt x_2,\wt x_3)=\\
&\left(
\begin{array}{cccc}
 -e^{-4 \Phi (t)} a_1^2 a_2^4 & 0 & 0 & 0 \\
 0 & \frac{a_2^2}{a_1^2 a_2^2+\wt x_2^2+\wt x_3^2} & \frac{\beta_{23} a_2^2 \wt x_3-\wt x_2}{a_1^2 a_2^2+\wt x_2^2+\wt x_3^2} & \frac{\beta_{23} \wt x_2 a_2^2+\wt x_3}{a_1^2 a_2^2+\wt x_2^2+\wt x_3^2} \\
 0 & \frac{\beta_{23} \wt x_3 a_2^2+\wt x_2}{a_1^2 a_2^2+\wt x_2^2+\wt x_3^2} & \frac{a_1^2 a_2^2+\left(\beta_{23}^2 a_2^4+1\right) \wt x_3^2}{a_2^2 \left(a_1^2 a_2^2+\wt x_2^2+\wt x_3^2\right)} & \frac{\left(\beta_{23}^2 a_2^4+1\right) \wt x_2 \wt x_3-\beta_{23} a_1^2 a_2^4}{a_2^2 \left(a_1^2 a_2^2+\wt x_2^2+\wt x_3^2\right)} \\
 0 & \frac{\beta_{23} a_2^2 \wt x_2-\wt x_3}{a_1^2 a_2^2+\wt x_2^2+\wt x_3^2} & \frac{\beta_{23} a_1^2 a_2^4+\left(\beta_{23}^2 a_2^4+1\right) \wt x_2 \wt x_3}{a_2^2 \left(a_1^2 a_2^2+\wt x_2^2+\wt x_3^2\right)} & \frac{a_1^2 a_2^2+\left(\beta_{23}^2 a_2^4+1\right) \wt x_2^2}{a_2^2 \left(a_1^2 a_2^2+\wt x_2^2+\wt x_3^2\right)} \\
\end{array}
\right)
\end{align*}
and with dilaton
$$
\wwt\Phi (t,\wt x_2,\wt x_3) = c_1 t-\frac{1}{2} \ln \left({a_2}(t)^2 \left({a_1}(t)^2 \,{a_2}(t)^2+{\wt x_2}^2+{\wt x_3}^2\right)\right)
$$
they satisfy \vbe. Tensors $\wwt\cf$ arising from $D_0\cdot I_B\cdot I_\beta$ and $D_0\cdot I_\beta\cdot I_B$ are too extensive to be displayed here. Nevertheless, it is straightforward to calculate them and verify that together with corresponding dilatons they satisfy \vbe.

\subsubsection{Factorized dualities}

\plti y \eqref{if1}, interpreted as
Buscher duality \wrt\ symmetry generated by $V_2$ and $V_3$, produces metric \eqref{mtzif6} with $\kappa = -1$ and functions ${a_2}(t)={a_3}(t)$ given by \eqref{B6-1_ai}. Dilaton calculated by the formula
\eqref{dualdilaton}
$$
\wh\Phi(t) = c_1 t - \frac{1}{2} \ln a_2(t)^4= -(c_1+p_2)\,t + const.$$
satisfies \vbe. Background obtained from $I_B \cdot I_F$ differs from this $\wh \cf$ only by a torsionless B-field that is the same as in \eqref{B6-1_B}.
Let us note that for $c_1=-p_2$ the metric is Ricci flat.

Dual \bkg\ produced by $D_0\cdot I_F$ reads
\begin{equation}\nonumber 
\wwt\cf(t,\wt x_2,\wt x_3)=\left(
\begin{array}{cccc}
 -e^{-4 \Phi (t)} a_1^2 a_2^4 & 0 & 0 & 0 \\
 0 & \frac{1}{\Delta} & -\frac{a_2^2 \wt x_2}{\Delta} & \frac{a_2^2 \wt x_3}{\Delta} \\
 0 & \frac{a_2^2 \wt x_2}{\Delta} & \frac{a_2^2 \left(a_1^2+a_2^2 \wt x_3^2\right)}{\Delta} & \frac{a_2^4 \wt x_2 \wt x_3}{\Delta} \\
 0 & -\frac{a_2^2 \wt x_3}{\Delta} & \frac{a_2^4 \wt x_2 \wt x_3}{\Delta} & \frac{a_2^2 \left(a_1^2+a_2^2 \wt x_2^2\right)}{\Delta} \\
\end{array}
\right)
\end{equation}
where
$$\Delta = a_1(t)^2+a_2(t)^2 \left(\wt x_2^2+\wt x_3^2\right).
$$
This \bkg\ and dilaton
$$\wwt \Phi(t,\wt x_2,\wt x_3)=c_1 t -\frac{1}{2}\ln\Delta
$$
satisfy \vbe. Background
\begin{align*}
&\wwt\cf(t,\wt y_2,\wt y_3)=\\
&\left(
\begin{array}{cccc}
 -e^{-4 \Phi (t)} a_1^2 a_2^4 & 0 & 0 & 0 \\
 0 & \frac{b_{23}^2 a_2^4+1}{\Delta} & -\frac{a_2^2 \left(b_{23} \wt y_3 a_2^2+\wt y_2\right)}{\Delta} & \frac{a_2^2 \wt y_3-b_{23} a_2^4 \wt y_2}{\Delta} \\
 0 & \frac{a_2^2 \wt y_2-b_{23} a_2^4 \wt y_3}{\Delta} & \frac{a_2^2 \left(a_1^2+a_2^2 \wt y_3^2\right)}{\Delta} & \frac{a_2^4 \left(b_{23} a_1^2+\wt y_2 \wt y_3\right)}{\Delta} \\
 0 & -\frac{a_2^2 \left(b_{23} \wt y_2 a_2^2+\wt y_3\right)}{\Delta} & -\frac{a_2^4 \left(b_{23} a_1^2-\wt y_2 \wt y_3\right)}{\Delta} & \frac{a_2^2 \left(a_1^2+a_2^2 \wt y_2^2\right)}{\Delta} \\
\end{array}
\right)
\end{align*}
where
$$
\Delta = \left(b_{23}^2 a_2(t)^4+1\right) a_1(t)^2+a_2(t)^2 \left(\wt y_2^2+\wt y_3^2\right)
$$
is obtained from $D_0\cdot I_B \cdot I_F$ and with dilaton
$$
\wwt \Phi(t,\wt y_2,\wt y_3) = c_1 t - \frac{1}{2}\ln \Delta
$$
it satisfies \vbe.

\section{Bianchi $II$ cosmology}\label{B_II}

Lie algebra $\mathfrak{d}= \mathfrak b_{II}\bowtie\ca$ of the
\dd\ $\D=({\B}_{II} |\A)$ is spanned by basis $(T_1,T_2,T_3,\wwt T^1,\wwt T^2,\wwt
T^3)$ where nontrivial \comrel s of $\mathfrak{b}_{II}$ are
\begin{equation}
\label{comB2} [ T_2, T_3]= T_1.
\end{equation}
Trace of structure constants is zero and group ${\B}_{II}$ is not
semisimple.

Cosmology invariant \wrt\ symmetry generated by left-invariant vector fields
\begin{equation}\nonumber 
V_1=\partial_{x_1},\qquad
V_2= -x_3\,\partial_{x_1}+\partial_{x_2},\qquad V_3=\partial_{x_3}
\end{equation}{satisfying \eqref{comB2} is given by the metric
\begin{align}\label{mtz21}
&\cf(t,x_2)=\nonumber \\
&\left(
\begin{array}{cccc}
-e^{-4 \Phi (t)}a_1(t)^2{a_2}(t)^2{a_3}(t)^2  & 0 & 0 & 0 \\
 0 &a_1(t)^2 & 0 & a_1(t)^2 x_2\\
 0 & 0 &{a_2}(t)^2 & 0 \\
 0 & a_1(t)^2 x_2 & 0 & a_1(t)^2 x_2^2 +{a_3}(t)^2 \\
\end{array}
\right)
\end{align}
where the functions $a_i(t)$ are
\begin{align}
{a_1}(t)&=e^{\Phi (t)} \sqrt{\frac{{p_1}}{ \cosh({p_1} t)}},\nonumber\\
{a_2}(t)&= e^{\Phi (t)+\frac{{p_2} t}{2}}\sqrt{\cosh ({p_1} t)},\label{B2_ai}\\
{a_3}(t)&= e^{\Phi (t)+\frac{{p_3} t}{2}}\sqrt{\cosh ({p_1}
t)}\nonumber
\end{align}
as in \cite{hokico}. For dilaton $\Phi(t)=c_1\, t$ the \vbe\ reduce
to
$$4c_1^2=p_3 p_2-p_1^2.$$
The background is torsionless and for $c_1 = 0$ also Ricci flat.

\subsection{\PL\ identities and dualities}

Unfortunately, we are not able to display general forms of matrices generating
\PL\ identities of Manin triple $(\cd,\mathfrak b_{II},\mathfrak a)$ 
because the expressions are too extensive. However, we were able to
decompose them into products of automorphisms, B-shifts and
$\beta$-shifts. To be more specific, all the solutions can be
written in one of the two forms
\begin{equation}\label{B2_I}
I_1=I_A \cdot I_B \cdot I_\beta, \qquad I_2=I_A \cdot I_\beta \cdot I_B
\end{equation}
where automorphisms have the form \eqref{IA} with
\begin{equation}\nonumber 
A=\left(
\begin{array}{ccc}
 \Lambda & 0 & 0 \\
 {a_{21}} & {a_{22}} & {a_{23}} \\
 {a_{31}} & {a_{32}} & {a_{33}} \\
\end{array}
\right),\quad \Lambda=\det \left(
\begin{array}{cc}
 {a_{22}} & {a_{23}} \\
 {a_{32}} & {a_{33}} \\
\end{array}
\right),
\end{equation}
B-shifts are generated by matrix \eqref{IB2}, and  $\beta$-shifts are given by
\begin{equation}\label{betashift}
I_\beta=\left(
\begin{array}{cccccc}
 1 & 0 & 0 & 0 & 0 & 0 \\
 0 & 1 & 0 & 0 & 0 & 0 \\
 0 & 0 & 1 & 0 & 0 & 0 \\
 0 & {\beta_{12}} & {\beta_{13}} & 1 & 0 & 0 \\
 -{\beta_{12}} & 0 & 0 & 0 & 1 & 0 \\
 -{\beta_{13}} & 0 & 0 & 0 & 0 & 1 \\
\end{array}
\right).
\end{equation}
There are no factorized dualities satisfying \eqref{C_str_const} and
\eqref{C_bilin}. \pltd ies can be obtained from identities by
left-multiplication  by the matrix \eqref{Delta} representing full
duality.

\subsection{Transformed \bkg s}

We already know that if \plti y decomposes as in \eqref{B2_I},
coordinate transformations can eliminate parameters of $I_A$ in the
resulting \bkg s. Thus it is sufficient to investigate the effects
of $I_B$, $I_\beta$ and their products.

\subsubsection{B-shifts}

Structure coefficients of Manin triple $(\cd,\mathfrak
b_{II},\mathfrak a)$ remain invariant under B-shift
\eqref{IB2} that transforms the \bkg\ \eqref{mtz21} to
\begin{align*}
&\wh\cf(t,x_2)=\\
&\left(
\begin{array}{cccc}
 -e^{-4 \Phi(t)}a_1^2{a_2}^2{a_3}^2 & 0 & 0 & 0 \\
 0 &a_1^2 & -{b_{12}} &a_1^2 x_2-{b_{13}} \\
 0 & {b_{12}} &{a_2}^2 & {b_{12}} x_2-{b_{23}} \\
 0 & x_2a_1^2+{b_{13}} & {b_{23}}-{b_{12}} x_2 &{a_3}^2+{a_1}^2 x_2^2 \\
\end{array}
\right).
\end{align*}
Up to gauge \tfn\ of the antisymmetric part it is equivalent to
\eqref{mtz21}. Together with dilaton $\wh \Phi(t)=c_1 t$ the \bkg\
satisfies \vbe.

{Dependence on $b_{23}$ can be eliminated in \bkg\ obtained from
$D_B = D_0\cdot I_B$ and we have
\begin{equation}\nonumber
\wwt\cf(t,y_1)=\left(
\begin{array}{cccc}
 -e^{-4 \Phi (t)} a_1^2 a_2^2 a_3^2 & 0 & 0 & 0 \\
 0 & \frac{a_2^2 a_3^2+\wt y_1^2}{\Delta} & \frac{b_{12} a_3^2-b_{13} \wt y_1}{\Delta} & \frac{b_{13} a_2^2+b_{12} \wt y_1}{\Delta} \\
 0 & -\frac{b_{12} a_3^2+b_{13} \wt y_1}{\Delta} & \frac{b_{13}^2+a_1^2 a_3^2}{\Delta} & \frac{a_1^2 \wt y_1-b_{12} b_{13}}{\Delta} \\
 0 & \frac{b_{12} \wt y_1-b_{13} a_2^2}{\Delta} & -\frac{\wt y_1 a_1^2+b_{12} b_{13}}{\Delta} & \frac{b_{12}^2+a_1^2 a_2^2}{\Delta} \\
\end{array}
\right)
\end{equation}
where
$$
\Delta={{a_1}(t)^2 \left({a_2}(t)^2{a_3}(t)^2+\wt y_1^2\right)+b_{13}^2 {a_2}(t)^2+b_{12}^2{a_3}(t)^2}.
$$
With dilaton
$$
\wwt\Phi(t,\wt y_1)=c_1  t-\frac{1}{2} \ln \Delta
$$
given by \eqref{dualdilaton} the \vbe\ are satisfied.
Results for the dual B-shift differ from the canonical dual obtained by $D_0$ not only by a shift in $y_1$ but also by other terms depending on $b_{12}, b_{13}$.}

\subsubsection{$\beta$-shifts}

Let us now investigate the transformation of metric \eqref{mtz21} given by  $\beta$-shift \eqref{betashift}.
This \PL\ identity generates {\begin{align*}
&\wh \cf (t,x_2)=\\
&\left(
\begin{array}{cccc}
 -e^{-4 \Phi (t)} a_1^2 a_2^2 a_3^2 & 0 & 0 & 0 \\
 0 & \frac{a_1^2}{\Delta} & \frac{a_1^2 a_2^2 \beta_{12}}{\Delta} & \frac{a_1^2 \left(\beta_{13} a_3^2+x_2\right)}{\Delta} \\
 0 & -\frac{a_1^2 a_2^2 \beta_{12}}{\Delta} & \frac{a_2^2 \left(a_1^2 a_3^2 \beta_{13}^2+1\right)}{\Delta} & -\frac{a_1^2 a_2^2 \beta_{12} \left(\beta_{13} a_3^2+x_2\right)}{\Delta} \\
 0 & \frac{a_1^2 \left(x_2-a_3^2 \beta_{13}\right)}{\Delta} & \frac{a_1^2 a_2^2 \beta_{12} \left(x_2-a_3^2 \beta_{13}\right)}{\Delta} & \frac{\left(a_1^2 a_2^2 \beta_{12}^2+1\right) a_3^2+a_1^2 x_2^2}{\Delta} \\
\end{array}
\right)
\end{align*}
where
$$
\Delta = \left(a_2(t)^2 \beta_{12}^2+a_3(t)^2 \beta_{13}^2\right) a_1(t)^2+1.
$$}
Together with corresponding  dilaton
\begin{equation}\nonumber
 \wh \Phi(t)=
 c_1  t-\frac{1}{2} \ln \Delta
\end{equation}
the \bkg\ satisfies \vbe. \plti y $I_B$ acting on this \bkg\ adds a
torsionless B-field, so we conclude that backgrounds obtained from
$I_B\cdot I_\beta$ and $I_\beta$ can be considered equivalent.
Despite the fact that $I_B$ and $I_\beta$ do not
commute, $\wh\cf$ obtained from $I_\beta \cdot I_B$ is exactly the
same as for $I_B\cdot I_\beta$.}

Dual \bkg\ {resulting from $D_\beta= D_0 \cdot I_\beta$ is
\begin{align*}
&\wwt \cf(t,\wt x_1)=\\
&\left(
\begin{array}{cccc}
 -e^{-4 \Phi (t)} a_1^2 a_2^2 a_3^2 & 0 & 0 & 0 \\
 0 & \frac{\left(a_3^2+\beta_{12}^2 a_1^2 \wt x_1^2\right) a_2^2+\left(\beta_{13}^2 a_1^2 a_3^2+1\right) \wt x_1^2}{a_1^2 \left(a_2^2 a_3^2+\wt x_1^2\right)} & \frac{a_3^2 \left(\beta_{13} \wt x_1-\beta_{12} a_2^2\right)}{a_2^2 a_3^2+\wt x_1^2} & -\frac{a_2^2 \left(\beta_{13} a_3^2+\beta_{12} \wt x_1\right)}{a_2^2 a_3^2+\wt x_1^2} \\
 0 & \frac{a_3^2 \left(\beta_{12} a_2^2+\beta_{13} \wt x_1\right)}{a_2^2 a_3^2+\wt x_1^2} & \frac{a_3^2}{a_2^2 a_3^2+\wt x_1^2} & \frac{\wt x_1}{a_2^2 a_3^2+\wt x_1^2} \\
 0 & \frac{a_2^2 \left(\beta_{13} a_3^2-\beta_{12} \wt x_1\right)}{a_2^2 a_3^2+\wt x_1^2} & -\frac{\wt x_1}{a_2^2 a_3^2+\wt x_1^2} & \frac{a_2^2}{a_2^2 a_3^2+\wt x_1^2} \\
\end{array}
\right)
\end{align*}}
{The dilaton is
\begin{equation}\nonumber
\wwt\Phi(t,\wt x_1)=c_1 t - \frac{1}{2}\ln\left(a_1(t)^2 \left(a_2(t)^2 a_3(t)^2+\wt x_1^2\right)\right)
\end{equation}
and it is interesting that it does not depend on $\beta_{12}$ and
$\beta_{13}$.} Together they satisfy \vbe.

Dual \bkg s and dilatons found from $D_0 \cdot I_B \cdot I_\beta$ and $D_0 \cdot I_\beta \cdot I_B$ are
too complicated to display and not particularly illuminating so we
omit them here. Nevertheless, one can check that they satisfy \vbe.

\section{Conclusions}

We have identified general forms of \PL {} identities and \pltd ies
for six-dimensional semi-Abelian Manin triples $\cd = \mathfrak  b\bowtie \ca$ where
$\mathfrak b$'s are Bianchi algebras that generate isometries of
Bianchi cosmologies. Subsequently we have  managed decomposing
both the
\PL\ identities and \pltd ies into simple factors, namely
automorphisms of Manin triples, B-shifts, $\beta$-shifts
and ``full'' or ``factorized'' dualities. This supports 
the conjecture posed in \cite{LuOst} that NATD group is generated by these elements. 
Finally, we have used these decompositions to transform Bianchi
cosmologies supplemented by dilaton fields
\cite{Batakis}. For these transformations we used \pltp y and dilaton formula described in Section 2.

In this way we have obtained  many new \bkg s and
corresponding dilatons that solve the Generalized
Supergravity Equations, and confirmed that
the Killing vector $\ci$ in Generalized
Supergravity Equations is given by trace of structure constants \cite{hokico}. One must, however, carefully evaluate what groups, more precisely what subgroups of \dd, truly
participate in the transformation since it influences resulting Killing vector. For factorized dualities these subgroups
often become Abelian and the Generalized Supergravity Equations
reduce to standard \vbe. New backgrounds, dilatons and corresponding Killing vectors are
summarized in the Tables in the Appendix. The backgrounds obtained by \PL\
identities are again invariant \wrt {} Bianchi groups. 

\newcommand{\mm}[2]{\begin{array}{c}#1,\\#2\\ \end{array}}
\newcommand{\mmm}[3]{\begin{array}{c}#1,\\#2,\\#3\\ \end{array}}

\section*{Appendix}

For reader's convenience we recapitulate \bkg s and dilatons yielded from \plti ies and dualities in the following tables. We add vector $\ci$ as well to indicate whether the \bkg s satisfy standard \vbe\ (in which case $\ci = 0$) or Generalized Supergravity Equations \eqref{betaG}--\eqref{betaPhi}. In the first column we display which one of the special transformations was used to get the result. Automorphisms $I_A$ are not mentioned. Nevertheless, since we want to include results obtained from general \PL\ identities and dualities, some parameters appearing in the tensors may arise from automorphisms. We recommend to check details in previous sections.

\begin{table}
\begin{center}
\scriptsize
{\renewcommand{\arraystretch}{1.4}
\begin{tabular}{|c || c |}
\hline
$\B_{II}$ & Transformed \bkg s, dilatons and vectors $\ci$   \\
\hline \hline
 $I_B$& $\wh\cf(t,x_2)=\left(
\begin{array}{cccc}
 -e^{-4 \Phi(t)}a_1^2 a_2^2 a_3^2 & 0 & 0 & 0 \\
 0 &a_1^2 & -{b_{12}} &a_1^2 x_2-{b_{13}} \\
 0 & {b_{12}} & a_2^2 & {b_{12}} x_2-{b_{23}} \\
 0 & x_2 a_1^2+{b_{13}} & {b_{23}}-{b_{12}} x_2 & a_3^2+a_1^2 x_2^2 \\
\end{array}
\right)$ \\
 & $ \wh \Phi(t)=c_1 t, \qquad \ci=0$\\
\hline
$D_B$ & $\wwt\cf(t,y_1)=\left(
\begin{array}{cccc}
 -e^{-4 \Phi (t)} a_1^2 a_2^2 a_3^2 & 0 & 0 & 0 \\
 0 & \frac{a_2^2 a_3^2+\wt y_1^2}{\Delta} & \frac{b_{12} a_3^2-b_{13} \wt y_1}{\Delta} & \frac{b_{13} a_2^2+b_{12} \wt y_1}{\Delta} \\
 0 & -\frac{b_{12} a_3^2+b_{13} \wt y_1}{\Delta} & \frac{b_{13}^2+a_1^2 a_3^2}{\Delta} & \frac{a_1^2 \wt y_1-b_{12} b_{13}}{\Delta} \\
 0 & \frac{b_{12} \wt y_1-b_{13} a_2^2}{\Delta} & -\frac{\wt y_1 a_1^2+b_{12} b_{13}}{\Delta} & \frac{b_{12}^2+a_1^2 a_2^2}{\Delta} \\
\end{array}
\right)$\\
 & $\Delta={{a_1}(t)^2 \left({a_2}(t)^2{a_3}(t)^2+\wt y_1^2\right)+b_{13}^2 {a_2}(t)^2+b_{12}^2{a_3}(t)^2}$\\
 & $ \wwt\Phi(t,\wt y_1)=c_1  t-\frac{1}{2} \ln \Delta, \qquad \mathcal{J}=0 $\\
\hline 
$I_\beta$ & $\wh \cf (t,x_2)=\left(
\begin{array}{cccc}
 -e^{-4 \Phi (t)} a_1^2 a_2^2 a_3^2 & 0 & 0 & 0 \\
 0 & \frac{a_1^2}{\Delta} & \frac{a_1^2 a_2^2 \beta_{12}}{\Delta} & \frac{a_1^2 \left(\beta_{13} a_3^2+x_2\right)}{\Delta} \\
 0 & -\frac{a_1^2 a_2^2 \beta_{12}}{\Delta} & \frac{a_2^2 \left(a_1^2 a_3^2 \beta_{13}^2+1\right)}{\Delta} & -\frac{a_1^2 a_2^2 \beta_{12} \left(\beta_{13} a_3^2+x_2\right)}{\Delta} \\
 0 & \frac{a_1^2 \left(x_2-a_3^2 \beta_{13}\right)}{\Delta} & \frac{a_1^2 a_2^2 \beta_{12} \left(x_2-a_3^2 \beta_{13}\right)}{\Delta} & \frac{\left(a_1^2 a_2^2 \beta_{12}^2+1\right) a_3^2+a_1^2 x_2^2}{\Delta} \\
\end{array}
\right)$\\
& $\Delta = \left(a_2(t)^2 \beta_{12}^2+a_3(t)^2 \beta_{13}^2\right) a_1(t)^2+1$\\
& $\wh \Phi(t)= c_1  t-\frac{1}{2} \ln \Delta, \qquad \ci=0$\\
\hline 
$D_\beta$ & $\begin{array}{l}
\wwt\cf(t,\wt x_1)=\\
\left(
\begin{array}{cccc}
 -e^{-4 \Phi (t)} a_1^2 a_2^2 a_3^2 & 0 & 0 & 0 \\
 0 & \frac{\left(a_3^2+\beta_{12}^2 a_1^2 \wt x_1^2\right) a_2^2+\left(\beta_{13}^2 a_1^2 a_3^2+1\right) \wt x_1^2}{a_1^2 \left(a_2^2 a_3^2+\wt x_1^2\right)} & \frac{a_3^2 \left(\beta_{13} \wt x_1-\beta_{12} a_2^2\right)}{a_2^2 a_3^2+\wt x_1^2} & -\frac{a_2^2 \left(\beta_{13} a_3^2+\beta_{12} \wt x_1\right)}{a_2^2 a_3^2+\wt x_1^2} \\
 0 & \frac{a_3^2 \left(\beta_{12} a_2^2+\beta_{13} \wt x_1\right)}{a_2^2 a_3^2+\wt x_1^2} & \frac{a_3^2}{a_2^2 a_3^2+\wt x_1^2} & \frac{\wt x_1}{a_2^2 a_3^2+\wt x_1^2} \\
 0 & \frac{a_2^2 \left(\beta_{13} a_3^2-\beta_{12} \wt x_1\right)}{a_2^2 a_3^2+\wt x_1^2} & -\frac{\wt x_1}{a_2^2 a_3^2+\wt x_1^2} & \frac{a_2^2}{a_2^2 a_3^2+\wt x_1^2} \\
\end{array}
\right)
\end{array} $\\
  & $\wwt\Phi(t,\wt x_1)=c_1 t - \frac{1}{2}\ln\left(a_1(t)^2 \left(a_2(t)^2 a_3(t)^2+\wt x_1^2\right)\right), \qquad \ci= 0$\\
\hline
\end{tabular}
\normalsize \caption{Results for \PL\ identities and dualities of Bianchi $II$ cosmology. Functions $a_i(t)$ are given by \eqref{B2_ai} and $\Phi(t)=c_1 t$. \label{tab:app_B2}}
}\end{center}
\end{table}

\begin{table}
\begin{center}
\scriptsize
{\renewcommand{\arraystretch}{1.4}
\begin{tabular}{|c || c |}
\hline
$\B_{III}$ & Transformed \bkg s, dilatons and vectors $\ci$   \\
\hline \hline
 $I_B$ & $\wh\cf (t,x_1) =\left(
\begin{array}{cccc}
 -1 & 0 & 0 & 0 \\
 0 & t^2 & -c_{15} & - c_{16} e^{-x_1} \\
 0 & c_{15} & 1 & 0 \\
 0 & c_{16} e^{-x_1} & 0 & t^2 e^{- 2 x_1} \\
\end{array}
\right)$ \\
 & $\wwt\Phi= 0 , \qquad \ci = 0$\\
\hline
$D_B$ & $\wwt\cf(t,\wt x_3)=\left(
\begin{array}{cccc}
 -1 & 0 & 0 & 0 \\
 0 & \frac{t^2}{t^4+c_{15}^2 t^2+\wt x_3^2} & \frac{t^2
   c_{15}}{t^4+c_{15}^2 t^2+\wt x_3^2} &
   -\frac{\wt x_3}{t^4+c_{15}^2 t^2+\wt x_3^2} \\
 0 & -\frac{t^2 c_{15}}{t^4+c_{15}^2 t^2+\wt x_3^2} &
   \frac{t^4+\wt x_3^2}{t^4+c_{15}^2 t^2+\wt x_3^2}
   & \frac{c_{15}\wt x_3}{t^4+c_{15}^2
   t^2+\wt x_3^2} \\
 0 & \frac{\wt x_3}{t^4+c_{15}^2 t^2+\wt x_3^2} & \frac{c_{15}\wt x_3}{t^4+c_{15}^2 t^2+\wt x_3^2} &
   \frac{t^2+c_{15}^2}{t^4+c_{15}^2 t^2+\wt x_3^2} \\
\end{array}
\right)$ \\
 & $\wwt\Phi(t,\wt x_3)=- \frac{1}{2}\ln\left(t^4+c_{15}^2t^2+\wt x_3^2\right), \qquad \ci = (0,-1,0,0)$ \\
\hline 
$I_{F_1}$ & $\wh\cf (t,x_1)=\left(
\begin{array}{cccc}
 -1 & 0 & 0 & 0 \\
 0 & t^2 & -c_{12} & - c_{13} e^{-x_1} \\
 0 & c_{12} & 1 & 0 \\
 0 & c_{13} e^{-x_1} & 0 & \frac{e^{-2 x_1}}{t^2} \\
\end{array}
\right)$ \\
& $\wh\Phi(t,x_1)= - \ln t - x_1 , \qquad \ci = 0$\\
\hline 
$D_{F_1}$ & $\wwt\cf(t,\wt x_3)=\left(
\begin{array}{cccc}
 -1 & 0 & 0 & 0 \\
 0 & \frac{1}{c_{15}^2+t^2 \left(\wt{x}_3^2+1\right)} & \frac{c_{15}}{c_{15}^2+t^2 \left(\wt{x}_3^2+1\right)} & -\frac{t^2 \wt{x}_3}{c_{15}^2+t^2 \left(\wt{x}_3^2+1\right)} \\
 0 & -\frac{c_{15}}{c_{15}^2+t^2 \left(\wt{x}_3^2+1\right)} & \frac{t^2 \left(\wt{x}_3^2+1\right)}{c_{15}^2+t^2 \left(\wt{x}_3^2+1\right)} & \frac{c_{15} t^2 \wt{x}_3}{c_{15}^2+t^2 \left(\wt{x}_3^2+1\right)} \\
 0 & \frac{t^2 \wt{x}_3}{c_{15}^2+t^2 \left(\wt{x}_3^2+1\right)} & \frac{c_{15} t^2 \wt{x}_3}{c_{15}^2+t^2 \left(\wt{x}_3^2+1\right)} & \frac{t^2 \left(c_{15}^2+t^2\right)}{c_{15}^2+t^2 \left(\wt{x}_3^2+1\right)} \\
\end{array}
\right)$ \\
  & $\wwt\Phi(t,\wt x_3)=-\frac{1}{2}\ln\left(c_{15}^2+t^2 \wt{x}_3^2+t^2\right), \qquad \ci = 0$ \\
\hline
\end{tabular}
\normalsize \caption{Results for \PL\ identities and dualities of Bianchi $III$ cosmology and dilaton $\Phi=0$. \label{tab:app_B3}}
}\end{center}
\end{table}

\begin{table}
\begin{center}
\scriptsize
{\renewcommand{\arraystretch}{1.4}
\begin{tabular}{|c || c |}
\hline
$\B_{V}$ & Transformed \bkg s, dilatons and vectors $\ci$   \\
\hline \hline
 $I_B$ & $\wh\cf (t,x_1) = \left(
\begin{array}{cccc}
 -1 & 0 & 0 & 0 \\
 0 & t^2 & -e^{x_1} c_{15} & -e^{x_1} c_{16} \\
 0 & e^{x_1} c_{15} & e^{2 x_1} t^2 & 0 \\
 0 & e^{x_1} c_{16} & 0 & e^{2 x_1} t^2 \\
\end{array}
\right)$ \\
 & $\wh\Phi= 0, \qquad \ci = 0$\\
\hline
$D_B$ & $\wwt\cf(t,\wt x_2,\wt x_3)=\left(
\begin{array}{cccc}
 -1 & 0 & 0 & 0 \\
 0 & \frac{t^2}{t^4+\tilde{x}_2^2+\tilde{x}_3^2} & \frac{\tilde{x}_2}{t^4+\tilde{x}_2^2+\tilde{x}_3^2} & \frac{\tilde{x}_3}{t^4+\tilde{x}_2^2+\tilde{x}_3^2} \\
 0 & -\frac{\tilde{x}_2}{t^4+\tilde{x}_2^2+\tilde{x}_3^2} & \frac{t^4+\tilde{x}_3^2}{t^2 \left(t^4+\tilde{x}_2^2+\tilde{x}_3^2\right)} & -\frac{\tilde{x}_2 \tilde{x}_3}{t^2 \left(t^4+\tilde{x}_2^2+\tilde{x}_3^2\right)} \\
 0 & -\frac{\tilde{x}_3}{t^4+\tilde{x}_2^2+\tilde{x}_3^2} & -\frac{\tilde{x}_2 \tilde{x}_3}{t^2 \left(t^4+\tilde{x}_2^2+\tilde{x}_3^2\right)} & \frac{t^4+\tilde{x}_2^2}{t^2 \left(t^4+\tilde{x}_2^2+\tilde{x}_3^2\right)} \\
\end{array}
\right)$ \\
 & $\wwt\Phi(t,\wt x_2,\wt x_3)= -\frac{1}{2} \ln \left(t^2 \left(x_2^2+x_3^2+t^4\right)\right), \qquad \ci = (0,2,0,0)$ \\
\hline 
$I_F$ & $\wh\cf (t,x_1)=\left(
\begin{array}{cccc}
 -1 & 0 & 0 & 0 \\
 0 & t^2 & -e^{x_1} c_{12} & -e^{x_1} c_{13} \\
 0 & e^{x_1} c_{12} & \frac{e^{2 x_1}}{t^2} & 0 \\
 0 & e^{x_1} c_{13} & 0 & \frac{e^{2 x_1}}{t^2} \\
\end{array}
\right)$ \\
& $\wh\Phi(t,x_1)= -2 \ln t+2 x_1, \qquad \ci = 0$\\
\hline 
$D_F$ & $\wwt\cf(t,\wt x_2,\wt x_3)= \left(
\begin{array}{cccc}
 -1 & 0 & 0 & 0 \\
 0 & \frac{1}{t^2 \left(\wt x_2^2+\wt x_3^2+1\right)} & \frac{\wt x_2}{\wt x_2^2+\wt x_3^2+1} & \frac{\wt x_3}{\wt x_2^2+\wt x_3^2+1} \\
 0 & -\frac{\wt x_2}{\wt x_2^2+\wt x_3^2+1} & \frac{t^2 \left(\wt x_3^2+1\right)}{\wt x_2^2+\wt x_3^2+1} & -\frac{t^2 \wt x_2 \wt x_3}{\wt x_2^2+\wt x_3^2+1} \\
 0 & -\frac{\wt x_3}{\wt x_2^2+\wt x_3^2+1} & -\frac{t^2 \wt x_2 \wt x_3}{\wt x_2^2+\wt x_3^2+1} & \frac{t^2 \left(\wt x_2^2+1\right)}{\wt x_2^2+\wt x_3^2+1} \\
\end{array}
\right)$ \\
  & $\wwt\Phi(t,\wt x_2,\wt x_3) = -\half\ln \left(
t^2 \left(x_2^2+x_3^2+1\right) \right), \qquad \ci = 0$ \\
\hline
\end{tabular}
\normalsize \caption{Results for \PL\ identities and dualities of Bianchi $V$ cosmology and dilaton $\Phi=0$. \label{tab:app_B5}}
}\end{center}
\end{table}

\begin{table}
\begin{center}
\scriptsize
{\renewcommand{\arraystretch}{1.4}
\begin{tabular}{|c || c |}
\hline
$\B_{VI_{\kappa}}$ & Transformed \bkg s, dilatons and vectors $\ci$   \\
\hline \hline
 $I_B$& $\wh\cf(t,x_1)=\left(
\begin{array}{cccc}
 -e^{-4 \Phi(t)} a_1^2 a_2^2 a_3^2 & 0 & 0 & 0 \\
 0 & a_1^2 & -e^{\kappa x_{1}} c_{15} & -e^{x_{1}} c_{16} \\
 0 & e^{\kappa x_{1}} c_{15} & e^{2 \kappa x_{1}} a_2^2 & 0 \\
 0 & e^{x_{1}} c_{16} & 0 & e^{2 x_{1}} a_3^2 \\
\end{array}
\right)$ \\
 & $ \wh \Phi (t)= c_1 t, \qquad \ci=0$\\
\hline
$D_B$ & $\wwt{\cf}(t,\wt x_2,\wt x_3)=\left(
\begin{array}{cccc}
 -e^{-4 \Phi (t)} a_1^2 a_2^2 a_3^2 & 0 & 0 & 0 \\
 0 & \frac{a_2^2 a_3^2}{\Delta} & \frac{\kappa a_3^2 \wt x_{2}}{\Delta} & \frac{a_2^2 \wt x_{3}}{\Delta} \\
 0 & -\frac{\kappa a_3^2 \wt x_{2}}{\Delta} & \frac{a_1^2 a_3^2+\wt x_{3}^2}{\Delta} & -\frac{\kappa \wt x_{2} \wt x_{3}}{\Delta} \\
 0 & -\frac{a_2^2 \wt x_{3}}{\Delta} & -\frac{\kappa \wt x_{2} \wt x_{3}}{\Delta} & \frac{a_1^2 a_2^2+\kappa^2 \wt x_{2}^2}{\Delta} \\
\end{array}
\right)$\\
 & $\Delta = a_1(t)^2 a_2(t)^2 a_3(t)^2+\kappa^2 \wt x_{2}^2 a_3(t)^2+a_2(t)^2 \wt x_{3}^2$\\
 & $ \wwt\Phi(t,\wt x_2, \wt x_3) = c_1 t - \frac{1}{2} \ln \Delta, \qquad \mathcal{J}=(0,\kappa+1,0,0) $\\
\hline 
$I_F$ & $ \wh\cf(t,x_1)=\left(
\begin{array}{cccc}
 -e^{-4 \Phi(t)}a_1^2 a_2^2 a_3^2 & 0 & 0 & 0 \\
 0 &a_1^2 & -e^{\kappa x_{1}} c_{12} & -e^{x_{1}} c_{13} \\
 0 & e^{\kappa x_{1}} c_{12} & \frac{e^{2 \kappa  x_1}}{a_2^2} & 0 \\
 0 & e^{\kappa x_{1}} c_{13} & 0 & \frac{e^{2 x_1}}{a_3^2} \\
\end{array}
\right)$\\
& $\wh\Phi(t,x_1)=c_1 t+(\kappa+1)x_1-\ln\left(a_2(t)a_3(t)\right),\qquad \ci=0$\\
\hline 
$D_F$ & $ \wwt\cf(t,\wt x_2, \wt x_3)=\left(
\begin{array}{cccc}
 -e^{-4 \Phi (t)} a_1^2 a_2^2 a_3^2 & 0 & 0 & 0 \\
 0 & \frac{1}{\Delta} & \frac{\kappa a_2^2 \wt x_{2}}{\Delta} & \frac{a_3^2 \wt x_{3}}{\Delta} \\
 0 & -\frac{\kappa a_2^2 \wt x_{2}}{\Delta} & \frac{a_2^2 \left(a_1^2+a_3^2 \wt x_{3}^2\right)}{\Delta} & -\frac{\kappa a_2^2 a_3^2 \wt x_{2} \wt x_{3}}{\Delta} \\
 0 & -\frac{a_3^2 \wt x_{3}}{\Delta} & -\frac{\kappa a_2^2 a_3^2 \wt x_{2} \wt x_{3}}{\Delta} & \frac{a_3^2 \left(a_1^2+\kappa^2 a_2^2 \wt x_{2}^2\right)}{\Delta} \\
\end{array}
\right)$\\
  & $\Delta = a_1(t)^2+\kappa^2 a_2(t)^2 \wt x_{2}^2+a_3(t)^2 \wt x_{3}^2$\\
  & $\wwt\Phi(t,\wt x_2, \wt x_3)=c_1 t - \frac{1}{2} \ln \Delta, \qquad \ci= 0$\\
\hline
\end{tabular}
\normalsize \caption{Results for \PL\ identities and dualities of Bianchi $VI_{\kappa}$ cosmology. Functions $a_i(t)$ are given by \eqref{B6k_ai}, $\kappa \neq -1$ and $\Phi(t)=c_1 t$. \label{tab:app_B6k}}
}\end{center}
\end{table}

\begin{table}
\begin{center}
\scriptsize
{\renewcommand{\arraystretch}{1.4}
\begin{tabular}{|c || c |}
\hline
$\B_{VI_{-1}}$ & Transformed \bkg s, dilatons and vectors $\ci$   \\
\hline \hline
 $I_B$& $\wh\cf(t,x_1)=\left(
\begin{array}{cccc}
 -e^{-4 \Phi(t)}a_1^2 a_2^4 & 0 & 0 & 0 \\
 0 & a_1^2 & -{b_{12}} e^{-x_1} & -{b_{13}} e^{x_1} \\
 0 & {b_{12}} e^{-x_1} & e^{-2 x_1} a_2^2 & -{b_{23}} \\
 0 & {b_{13}} e^{x_1} & {b_{23}} & e^{2 x_1} a_2^2 \\
\end{array}
\right)$ \\
 & $ \wh\Phi=c_1 t, \qquad \ci=0$\\
\hline
$D_B$ & $\wwt\cf(t, \wt y_2, \wt y_3)=\left(
\begin{array}{cccc}
 -e^{-4 \Phi (t)} a_1^2 a_2^4 & 0 & 0 & 0 \\
 0 & \frac{a_2^4+b_{23}^2}{\Delta} & -\frac{\wt y_2 a_2^2+b_{23} \wt y_3}{\Delta} & \frac{a_2^2 \wt y_3-b_{23} \wt y_2}{\Delta} \\
 0 & \frac{a_2^2 \wt y_2-b_{23} \wt y_3}{\Delta} & \frac{a_1^2 a_2^2+\wt y_3^2}{\Delta} & \frac{b_{23} a_1^2+\wt y_2 \wt y_3}{\Delta} \\
 0 & -\frac{\wt y_3 a_2^2+b_{23} \wt y_2}{\Delta} & \frac{\wt y_2 \wt y_3-b_{23} a_1^2}{\Delta} & \frac{a_1^2 a_2^2+\wt y_2^2}{\Delta} \\
\end{array}
\right)$\\
 & $\Delta=\left(a_2(t)^4+b_{23}^2\right) a_1(t)^2+a_2(t)^2 \left(\wt y_2^2+\wt y_3^2\right)$\\
 & $\wwt\Phi(t, \wt y_2, \wt y_3)=c_1  t-\frac{1}{2} \ln \Delta, \qquad \ci=0$\\
\hline 
$I_\beta$ & $\wh\cf(t,x_1)=\left(
\begin{array}{cccc}
 -e^{-4 \Phi(t)}a_1^2 a_2^4 & 0 & 0 & 0 \\
 0 & a_1^2 & 0 & 0 \\
 0 & 0 & \frac{e^{-2 x_1} a_2^2}{\beta_{23}^2 a_2^4+1} & \frac{{\beta_{23}}
  a_2^4}{\beta_{23}^2 a_2^4+1} \\
 0 & 0 & -\frac{{\beta_{23}} a_2^4}{\beta_{23}^2 a_2^4+1} & \frac{e^{2 x_1}
  a_2^2}{\beta_{23}^2 a_2^4+1} \\
\end{array}
\right)$\\
& $ \wh\Phi(t)=c_1 t - \frac{1}{2} \ln \left(\beta_{23}^2 a_2(t)^4+1 \right),\qquad \ci=0$\\
\hline 
$D_\beta$ & $\begin{array}{l} \wwt\cf(t,\wt x_2,\wt x_3)=\\
\left(
\begin{array}{cccc}
 -e^{-4 \Phi (t)} a_1^2 a_2^4 & 0 & 0 & 0 \\
 0 & \frac{a_2^2}{a_1^2 a_2^2+\wt x_2^2+\wt x_3^2} & \frac{\beta_{23} a_2^2 \wt x_3-\wt x_2}{a_1^2 a_2^2+\wt x_2^2+\wt x_3^2} & \frac{\beta_{23} \wt x_2 a_2^2+\wt x_3}{a_1^2 a_2^2+\wt x_2^2+\wt x_3^2} \\
 0 & \frac{\beta_{23} \wt x_3 a_2^2+\wt x_2}{a_1^2 a_2^2+\wt x_2^2+\wt x_3^2} & \frac{a_1^2 a_2^2+\left(\beta_{23}^2 a_2^4+1\right) \wt x_3^2}{a_2^2 \left(a_1^2 a_2^2+\wt x_2^2+\wt x_3^2\right)} & \frac{\left(\beta_{23}^2 a_2^4+1\right) \wt x_2 \wt x_3-\beta_{23} a_1^2 a_2^4}{a_2^2 \left(a_1^2 a_2^2+\wt x_2^2+\wt x_3^2\right)} \\
 0 & \frac{\beta_{23} a_2^2 \wt x_2-\wt x_3}{a_1^2 a_2^2+\wt x_2^2+\wt x_3^2} & \frac{\beta_{23} a_1^2 a_2^4+\left(\beta_{23}^2 a_2^4+1\right) \wt x_2 \wt x_3}{a_2^2 \left(a_1^2 a_2^2+\wt x_2^2+\wt x_3^2\right)} & \frac{a_1^2 a_2^2+\left(\beta_{23}^2 a_2^4+1\right) \wt x_2^2}{a_2^2 \left(a_1^2 a_2^2+\wt x_2^2+\wt x_3^2\right)} \\
\end{array}
\right)
\end{array}$\\
 & $\wwt\Phi (t,\wt x_2,\wt x_3) = c_1 t-\frac{1}{2} \ln \left({a_2}(t)^2 \left({a_1}(t)^2 \,{a_2}(t)^2+{\wt x_2}^2+{\wt x_3}^2\right)\right), \qquad \ci= 0$\\
\hline 
$I_F$ & $\wh\cf(t,x_1)=\left(
\begin{array}{cccc}
 -e^{-4 \Phi(t)} a_1^2 a_2^4 & 0 & 0 & 0 \\
 0 & a_1^2 & 0 & 0 \\
 0 & 0 & \frac{e^{-2 x_1}}{a_2^2} & 0 \\
 0 & 0 & 0 & \frac{e^{2 x_1}}{a_2^2} \\
\end{array}
\right)$\\
& $\wh\Phi(t) = c_1 t - \frac{1}{2} \ln a_2(t)^4,\qquad \ci=0$\\
\hline 
$D_F$ & $\wwt\cf(t,\wt x_2,\wt x_3)=\left(
\begin{array}{cccc}
 -e^{-4 \Phi (t)} a_1^2 a_2^4 & 0 & 0 & 0 \\
 0 & \frac{1}{\Delta} & -\frac{a_2^2 \wt x_2}{\Delta} & \frac{a_2^2 \wt x_3}{\Delta} \\
 0 & \frac{a_2^2 \wt x_2}{\Delta} & \frac{a_2^2 \left(a_1^2+a_2^2 \wt x_3^2\right)}{\Delta} & \frac{a_2^4 \wt x_2 \wt x_3}{\Delta} \\
 0 & -\frac{a_2^2 \wt x_3}{\Delta} & \frac{a_2^4 \wt x_2 \wt x_3}{\Delta} & \frac{a_2^2 \left(a_1^2+a_2^2 \wt x_2^2\right)}{\Delta} \\
\end{array}
\right)$\\
  & $\Delta = a_1(t)^2+a_2(t)^2 \left(\wt x_2^2+\wt x_3^2\right)$\\
  & $\wwt \Phi(t,\wt x_2,\wt x_3)=c_1 t -\frac{1}{2}\ln\Delta, \qquad \ci= 0$\\
\hline
\end{tabular}
\normalsize \caption{Results for \PL\ identities and dualities of Bianchi $VI_{-1}$ cosmology. Functions $a_i(t)$ are given by \eqref{B6-1_ai} and $\Phi(t)=c_1 t$. \label{tab:app_B6-1}}
}\end{center}
\end{table}

\end{document}